\documentclass[english, reprint, aps, pra, showpacs]{revtex4-1}
\usepackage[T1]{fontenc}
\usepackage[latin9]{inputenc}
\usepackage{amsmath}
\usepackage{amssymb}
\usepackage{graphicx}
\usepackage{babel}
\begin{document}
\global\long\def\ket#1{|#1\rangle}
\global\long\def\bra#1{\langle#1|}
\global\long\def\op#1{\hat{\mathsf{#1}}}
\global\long\def\mean#1{\overline{#1}}
\global\long\def\k{\kappa}
\global\long\def\i{\mathrm{\mathrm{i}}}
\global\long\def\ind#1{\mathrm{#1}}
\global\long\def\vek#1{\bm{#1}}
\global\long\def\Dot#1{\frac{\mathrm{d}}{\mathrm{d}t}#1}

\title{Unraveling Quantum Brownian Motion: Pointer States and their Classical
Trajectories}

\author{Lutz Sörgel and Klaus Hornberger}

\affiliation{Faculty of Physics, University of Duisburg-Essen, Lotharstraße 1,
47048 Duisburg, Germany}

\date{\today}


\begin{abstract}
We characterize the pointer states generated by the master equation
of quantum Brownian motion and derive stochastic equations for the
dynamics of their trajectories in phase space. Our method is based
on a Poissonian unraveling of the master equation whose deterministic
part exhibits soliton-like solutions that can be identified with the
pointer states. In the semiclassical limit, their phase space trajectories
turn into those of classical diffusion, yielding a clear picture of
the induced quantum-classical transition. 
\end{abstract}
\maketitle

\section{Introduction}

Decoherence through environmental coupling is a crucial ingredient
for understanding the quantum-to-classical transition \cite{Joos_Deco_App_Class_World,Zurek_Review_Dec_Eins_Quant_Origin_Class,MSchlosshBook}.
As a prototypical problem, one may consider the motion of a quantum
particle interacting with an ideal gas environment. The interaction
with the environment selects a preferred set of states, the so-called
pointer states \cite{Zurek_What_Mixtures_PS,Zurek_CohStatesViaDecoh1993,DiosiKieferRobustnessDiffPS}.
These special states of the particle remain pure for a relatively
long time, whereas their superpositions decohere very fast into a
mixture. For simple master equations the pointer states are known
\cite{Zurek_CohStatesViaDecoh1993,DiosiKieferRobustnessDiffPS,Eisert2004_ExactDecoToPS,Busse_Hornberger_PS_Nonpertubative_Environment,Busse_Hornberger_PS_CollDeco}
to be localized wave packets moving in phase space. In the semiclassical
limit they turn into arbitrarily sharp peaks moving along classical
phase space trajectories. As such, pointer states are ideally suited
to explain the emergence of classicality within a quantum framework.

The name \emph{pointer states }derives from the fact that their characteristics
matches the behavior of a pointer of a measuring apparatus. Namely,
if a quantum system is in an eigenstate of the observable we measure
with an apparatus its pointer will turn to a specific value on the
dial and stay there. On the other hand, if the system is in a superposition
of two nondegenerate eigenstates one will not find the pointer of
the apparatus to be in a superposition of the two corresponding values,
as prescribed by unitarity, but pointing to either one or the other
position with probabilities determined by Born's rule.

In this article, we identify the pointer states of the master equation
of quantum Brownian motion (QBM) and derive their equations of motion.
Specifically, we understand as QBM the dissipative dynamics of a free
particle resulting from many small environmental collisions as described
by a Lindblad type master equation \cite{Lindblad1976_OnTheGenerators,GoriniKossakowskiSudarshan1976}.
One possibility to obtain this master equation is to consider a particle
linearly coupled to a bath of harmonic oscillators \cite{Caldeira_Leggett_PathIntegralQBM,UnruhZurek1989_ReductionWavePacketInQBM,StrunzHaakeBraun2003_UnivDecForMacroSuperpos,HaenggiIngold2005_FundamentalAspectsQBM}
and to take the high-temperature limit of the reduced particle dynamics;
one arrives at the master equation of the Caldeira-Leggett model \cite{Caldeira_Leggett_PathIntegralQBM,Zurek_CohStatesViaDecoh1993,StrunzDiosiGisin1999_QTForBM}.
To get a Lindblad master equation an additional term has to be appended
\cite{Diosi1993_HighTempMarkovEqForQBM,PetruccioneVacchini2005_QDescrOfBrownianMotion,Vacchini_Hornberger_RelaxationDyn_QBM}
making the equation completely positive. Alternatively, the Lindblad
master equation of QBM can be obtained by considering a dissipative
limit of the general particle motion in a gaseous environment as governed
by the so-called quantum linear Boltzmann equation \cite{Vacchini2000_CompPosQuDiss,Hornberger2006_MEQUParticleGas,Vacchini_Hornberger_RelaxationDyn_QBM,Vacchini_Hornberger_QLBE_Review}.

By considering quantum trajectories using a particular stochastic
process in Hilbert space, the so-called pointer state unraveling,
we find the pointer states of QBM to be rotated Gaussians in phase
space, in line with the encounter of Gaussian pointer states in other
models of linear environmental coupling \cite{Eisert2004_ExactDecoToPS,DiosiKieferRobustnessDiffPS}.
Moreover, we show how superpositions of separated wave packets decohere
into mixtures of pointer states with weights given by the absolute
value of the initial amplitude in the superposition (i.e. Born's rule).
Finally, we derive the stochastic equations of motion for the location
of the pointer states in phase space and we show that they turn into
the equation of classical diffusion in the semiclassical limit, thus
providing a comprehensive picture of the quantum-to-classical transition
for the important dissipative process of QBM. For times of evolution
much greater than the characteristic decoherence time, it is then
justified to approximate the dynamics by a mixture of pointer states
moving on trajectories in phase space, instead of calculating the
whole evolution of the density matrix.

Our method of using the pointer-state unraveling to identify the pointer
basis and to provide a perspective on the quantum-to-classical transition
for particle motion has already been demonstrated for the non-dissipative
case of pure collisional decoherence \cite{GallisFleming1990,HornbergerSipe2003_CollDecoReexamined,Busse_Hornberger_PS_CollDeco}.
In that case, the pointer states are exponentially localized, rather
than Gaussian, and the emerging classical equations of motion are
Newton's equations. In the present case of dissipative dynamics we
derive a stochastic equation of motion for the pointer states in order
to get the Langevin equation corresponding to classical diffusion.
This requires a subtle analysis of the dynamics of the quantum trajectories.

The paper is organized as follows: At first, in section \ref{sec:Pointer-States},
we give a general definition of pointer states. In the following section
\ref{sec:Quantum-Brownian-Motion} we introduce and briefly discuss
the Lindblad master equation of QBM and introduce its pointer state
unraveling. The careful examination of this particular quantum trajectory
approach is the topic of the remainder of the paper: The deterministic
part of the unraveling is analyzed in section \ref{sec:Pointer-states-of QBM}
providing us with the Gaussian pointer states of QBM as well as with
the deterministic part of the classical equations of motion. By looking
at the full unraveling, i.e. the deterministic and stochastic part,
we show in section \ref{sec:Unraveling-Quantum-Brownian} that any
superposition of separated wave packets turns into a mixture of pointer
states with the expected probabilities. In addition, we identify the
trajectories of the pointer states; they exhibit a stochastic motion
in phase space and yield diffusion in the semiclassical limit. Section
\ref{sec:Conclusion} contains our conclusions.

\section{\label{sec:Pointer-States}Pointer states}

The evolution of a Markovian open quantum system can be described
by a Lindblad master equation \cite{GoriniKossakowskiSudarshan1976,Lindblad1976_OnTheGenerators,Breuer_Petruccione_OpenQS}.
For this equation to exhibit pointer states we expect it to display
a time scale separation into a short decoherence time $t_{\ind{dec}}$
and a much longer dynamical time scale which has also a classical
meaning. The set of pointer states is characterized by the property
that superpositions of mutually orthogonal pointer states decay into
a mixture on the decoherence time scale whereas a single pointer state
remains relatively stable and changes on the longer time scale only.

Specifically, for a superposition of pointer states
\begin{equation}
\ket{\psi_{0}}=\sum_{\alpha=1}^{N}c_{\alpha}(0)\ket{\pi_{\alpha}(0)},\label{eq:initial state ps}
\end{equation}
the solution of the master equation for times much greater than $t_{\mathrm{dec}}$
is well approximated by the mixture of pointer states \cite{Breuer_Petruccione_OpenQS}
\begin{equation}
\rho_{0}=\ket{\psi_{0}}\bra{\psi_{0}}\rightarrow\rho_{t}\simeq\sum_{\alpha=1}^{N}\,\mathrm{Prob(\alpha|\rho_{0})\ket{\pi_{\alpha}(t)}}\bra{\pi_{\alpha}(t)}.\label{eq:sol_of_ME_by_ps}
\end{equation}
A proper set of pointer states forms a basis, is independent of the
initial state, and depends only on the environment and the coupling
operators. As a consequence and according to Born's rule, the only
dependence of the mixture (\ref{eq:sol_of_ME_by_ps}) on the initial
state $\rho_{0}$ is via the probabilities
\begin{equation}
\mathrm{Prob}(\alpha|\rho_{0})=\mathrm{Tr}\left(\rho_{0}\ket{\pi_{\alpha}(0)}\bra{\pi_{\alpha}(0)}\right)=|c_{\alpha}(0)|^{2}.\label{eq:prob_for_ps}
\end{equation}
The time dependence of the pointer states $\ket{\psi_{\alpha}(t)}$
occurs on the longer time scale only. Once the pointer states and
their time dependence are characterized, the dynamics of the master
equation is fully captured by Eq.~(\ref{eq:sol_of_ME_by_ps}) for
times greater than $t_{\ind{dec}}$.

There are different possibilities to identify the pointer basis. A
well-known one requires that the pointer states should be the least
entropy producing states of the system; this is the so called predictability
sieve \cite{Zurek1993_PredictSieve,Zurek_predSieve_Ps}. Here, we
follow another approach, in which the pointer states are identified
as soliton-like solutions of a nonlinear equation associated with
the master equation. This nonlinear pointer state equation (NLPSE)
can be obtained heuristically by identifying the pure state evolution
closest to the master equation evolution \cite{Diosi1986_StochasticPureStateRep,Gisin_Rigo_Nonlin_Ses}.

The pointer state unraveling, a piecewise deterministic stochastic
process in Hilbert space, connects the solution of the master equation
with the NLPSE, since the latter describes the deterministic part
of the unraveling. In the next section, we present the master equation
of QBM together with its pointer state unraveling.

\section{\label{sec:Quantum-Brownian-Motion}Quantum Brownian motion}

\subsection{Master equation of quantum Brownian motion}

The master equation of QBM describes the quantum state of a massive
marker particle linearly coupled to a high-temperature gas environment
and experiencing decoherence as well as dissipation. For the sake
of simplicity, we restrict ourselves here to the one dimensional case
and use dimensionless variables by introducing time, length and momentum
scales
\begin{equation}
\mathrm{T}=\frac{1}{2\gamma},\quad\mathrm{L}=\frac{1}{2\gamma}\sqrt{\frac{k_{\mathrm{B}}T_{\ind{env}}}{m}},\quad\mathrm{P}=\sqrt{mk_{\mathrm{B}}T_{\ind{env}}}.\label{eq:scales}
\end{equation}
They involve the mass of the particle $m$, the temperature of the
surrounding environment $T_{\ind{env}}$, and the friction constant
$\gamma$. These scales do not depend on $\hbar$ and are therefore
convinient in describing the semiclassical limit of QBM. In these
units the canonical commutator takes the form $\left[\op x,\op p\right]=\i/\k$
and the momentum operator in position representation is $\bra x\op p\ket{\Phi}=-\i/\k\,\partial_{x}\Phi(x)$,
where we introduced the dimensionless parameter
\begin{equation}
\k=\frac{k_{\mathrm{B}}T_{\ind{env}}}{2\gamma\hbar}.\label{eq:k equals}
\end{equation}
The semiclassical limit is then obtained by sending $\k\rightarrow\infty$.

The Lindblad master equation of QBM is obtained by using a single
Lindblad operator
\begin{equation}
\op L=\sqrt{2}\left(\k\op x+\frac{\i}{4}\op p\right)\label{eq:Lindblad_QBM}
\end{equation}
and the Hamiltonian
\begin{equation}
\op H=\hbar\k\left(\frac{\op p^{2}}{2}+\frac{1}{4}\left\{ \op x,\op p\right\} \right).\label{eq:Hamiltonian_QBM}
\end{equation}
The anti-commutator $\left\{ \op x,\op p\right\} /4$ describes a
rescaling of the particle energy due to the environmental coupling.
Inserting (\ref{eq:Lindblad_QBM}) and (\ref{eq:Hamiltonian_QBM})
into the general Lindblad master equation \cite{Lindblad1976_OnTheGenerators,Breuer_Petruccione_OpenQS}
\begin{equation}
\Dot{\rho}=\frac{1}{\i\hbar}\left[\op H,\rho\right]+\left(\op L\rho\op L^{\dagger}-\frac{1}{2}\op L^{\dagger}\op L\rho-\frac{1}{2}\rho\op L^{\dagger}\op L\right),\label{eq:general Lindblad ME}
\end{equation}
we arrive at the Lindblad master equation of QBM \cite{Vacchini_Hornberger_RelaxationDyn_QBM,Vacchini_Hornberger_QLBE_Review}
\begin{align}
\Dot{\rho} & =-\i\frac{\k}{2}\left[\op p^{2},\rho\right]-\i\frac{\k}{2}\left[\op x,\left\{ \op p,\rho\right\} \right]-\k^{2}\left[\op x,\left[\op x,\rho\right]\right]\nonumber \\
 & \hspace{1em}-\frac{1}{16}\left[\op p,\left[\op p,\rho\right]\right].\label{eq:ME QBM}
\end{align}

Equation (\ref{eq:ME QBM}) is the Lindblad generalization of the
famous Caldeira-Leggett master equation \cite{Caldeira_Leggett_PathIntegralQBM,Diosi1993_HighTempMarkovEqForQBM,Breuer_Petruccione_OpenQS},
having in common the first three terms on the right-hand side. They
describe free motion, friction, and momentum diffusion, respectively.
The last term of (\ref{eq:ME QBM}), which ensures complete positivity,
can be identified with a position diffusion. This position diffusion
term is not present in the Caldeira-Leggett master equation, and its
influence decreases with growing $\k$.

When writing (\ref{eq:ME QBM}) in the Wigner phase space representation
one can draw the semiclassical limit, $\k\rightarrow\infty$, to see
the formal analogy between the evolution equation for the Wigner function
and the Fokker-Planck equation for the classical phase space distribution
of classical diffusion \cite{Caldeira_Leggett_PathIntegralQBM,Vacchini_Hornberger_RelaxationDyn_QBM}.
Further analogies to the classical case of Brownian motion can be
seen if one considers the mean values and variances of the position
and momentum operators \cite{Breuer_Petruccione_OpenQS}. Most prominently,
one finds that the mean of the momentum operator is exponentially
damped and that the variance of the position operator grows linearly
for asymptotically large times; both properties are characteristics
of classical Brownian motion.

\subsection{\label{sub:unraveling QBM}Unraveling of the master equation}

In order to derive the pointer states of QBM and their motion in phase
space, we replace the deterministic master equation evolution for
the density matrix (\ref{eq:ME QBM}) by an equivalent stochastic
pure state evolution in Hilbert space, the pointer state unraveling.
The equivalence of an unraveling to a master equation is provided
by the fact that the ensemble average over the quantum trajetories
yields the solution of the master equation \cite{Diosi1986_StochasticPureStateRep,Gardiner1992_QSDEMethods,DalibardCastinMolmer1992_WaveFunctionApprInQO,Molmer1993_MonteCarloMethodsQO,Gisin_Rigo_Nonlin_Ses,RigoGisin1996_UnravMEEmergenceClass,Busse_Hornberger_PS_CollDeco}.
Since the decomposition of a density matrix into a mixture of pure
states is not unique, many ensembles of quantum trajectories will
recover the same solution of the master equation, hence the choice
for an unraveling is ambiguous. One distinguishes between continous
unravelings based on a Wiener process \cite{Gisin1984_QuMeasStochProc,Diosi1988_QuantStochProcModelStateVectorRed,GhirardiPearleRimini1990_MarkovProcessesHilbertSpaceandCSL,GisinPercival1992_QSDAppliedToOpenSys,StrunzDiosiGisin1999_QTForBM},
where \emph{quantum state diffusion} is the most prominent, and piecewise
deterministic unravelings \cite{Diosi1986_StochasticPureStateRep,DalibardCastinMolmer1992_WaveFunctionApprInQO,Molmer1993_MonteCarloMethodsQO,Gisin_Rigo_Nonlin_Ses,Busse_Hornberger_PS_CollDeco,Busse_Hornberger_PS_Nonpertubative_Environment}
based on a Poisson process.

We will use here a piecewise deterministic unraveling which was already
employed in the case of collisional decoherence to determine the pointer
states and their dynamics \cite{Busse_Hornberger_PS_CollDeco,Busse_Hornberger_PS_Nonpertubative_Environment,RigoGisin1996_UnravMEEmergenceClass}.
In our case of QBM the deterministic part is formed by the nonlinear
operator \cite{Gisin_Rigo_Nonlin_Ses}
\begin{align}
\op N[\psi]\ket{\psi}= & \Bigg(-\i\frac{\k}{2}\left(\op p^{2}-\mean{p^{2}}\right)\nonumber \\
 & \hspace{1em}+\k^{2}\left(V_{x}-\left(\op x-\mean x\right)^{2}\right)+\frac{1}{16}\left(V_{p}-\left(\op p-\mean p\right)^{2}\right)\nonumber \\
 & \hspace{1em}+\frac{\k}{4}\left(\i C_{xp}-\frac{1}{\k}\right)-\i\frac{\k}{2}\left(\op x-\mean x\right)\left(\op p+\mean p\right)\Bigg)\ket{\psi},\label{eq:deterministic part QBM}
\end{align}
with expectation values denoted as $\mean A=\langle\op A\rangle=\bra{\psi}\op A\ket{\psi},$
variances $V_{A}=\mean{A^{2}}-\mean A^{2}$ and the covariance $C_{xp}=\mean{\left\{ x,p\right\} }-2\mean x\,\mean p$.
Due to the time dependence of the state $\ket{\psi}$, all expectation
values are time dependent. The time variable is omitted here for the
sake of readability.

The evolution equation
\begin{equation}
\Dot{\ket{\psi}}=\op N[\psi]\ket{\psi},\label{eq:NLPSE QBM}
\end{equation}
is called the nonlinear pointer state equation (NLPSE) because it
is expected to exhibit soliton-like solutions which we will identify
as the pointer states. Soliton-like is understood here in the sense
that the shape of the wave function is time-independent in phase space.
As mentioned above, the NLPSE (\ref{eq:NLPSE QBM}) can also be obtained
by looking for the particular pure state equation that mimics the
master equation most closely \cite{Gisin_Rigo_Nonlin_Ses,DiosiKieferRobustnessDiffPS}.

To obtain the whole unraveling a stochastic jump part must be added
to the deterministic NLPSE (\ref{eq:NLPSE QBM}) yielding 
\begin{align}
\ket{\mathrm{d}\psi} & =\op N[\psi]\ket{\psi}\mathrm{d}t+\left(\frac{\op J[\psi]}{\|\op J[\psi]\ket{\psi}\|}-1\right)\ket{\psi}\mathrm{d}N,\label{eq:ps unraveling QBM}
\end{align}
with the nonlinear jump operator
\begin{align}
\op J[\psi] & =\sqrt{2}\left(\k\left(\op x-\mean x\right)+\frac{\i}{4}\left(\op p-\mean p\right)\right).\label{eq:jump op QBM}
\end{align}
The stochastic Poisson increment $\mathrm{d}N$ can take values $\mathrm{d}N\in\{0,1\}$
and satisfies the Poisson increment rule 
\begin{equation}
\mathrm{d}N^{2}=\mathrm{d}N.\label{eq:Poisson increment rule}
\end{equation}
The nonlinearity of the jump operator (\ref{eq:jump op QBM}) arrises
because of its dependence on the position and momentum expectation
values which depend on the state $\ket{\psi}$. To fully define the
above Poisson process we have to specify the ensemble average of the
Poisson increment. It is determined by the rate $r=\bra{\psi}\op J^{\dagger}[\psi]\op J[\psi]\ket{\psi}$
of the process, given in our case by

\begin{align}
\frac{\mathrm{E}\left[\mathrm{d}N\right]}{\mathrm{d}t} & =r=2\k^{2}V_{x}+\frac{1}{8}V_{p}-\frac{1}{2},\label{eq:jump rate QBM}
\end{align}
with $\mathrm{E}\left[\cdot\right]$ denoting the ensemble average.

A straightforward calculation shows that this unraveling indeed recovers
the master equation (\ref{eq:ME QBM}) in the ensemble average \cite{Gisin_Rigo_Nonlin_Ses}.
It is also straightforward to verify that the jump rate in this unraveling
is proportional to the change of purity of the state
\begin{equation}
r=\frac{1}{2}\Dot{\mathrm{Tr}\left(1-\left(\ket{\psi}\bra{\psi}\right)^{2}\right)}.\label{eq: QBM tot jump rate}
\end{equation}
This finding fits with the expected behavior of a pointer state unraveling:
Once the pointer state has been reached due to the deterministic part
of the unraveling (as determined by the NLPSE) it should be relatively
stable. This implies that the change of purity of the pointer state
is small and so is therefore the total jump rate (\ref{eq: QBM tot jump rate}).
As a consequence, once a pointer state is formed in a quantum trajectory
of this unraveling, it remains a pointer state as long as possible.

\section{\label{sec:Pointer-states-of QBM}Pointer states of quantum Brownian
motion}

In this section we present the pointer states of QBM as soliton-like
solutions of the NLPSE (\ref{eq:NLPSE QBM}). After that we analyze
the action of the NLPSE on a superposition of separated wave packets,
which will help us to obtain Born's rule in section \ref{sec:Unraveling-Quantum-Brownian}.

\subsection{Single wave packets\label{sub:NLPSE Single Wave Packets}}

A lengthy but straightforward calculation confirms that the NLPSE
(\ref{eq:NLPSE QBM}) is solved by a Gaussian state of the form
\begin{align}
\psi(x) & =\frac{1}{\left(2\pi V_{x}\right)^{1/4}}\exp\Bigg(-\frac{\left(x-\mean x\right)^{2}}{4V_{x}}\left(1-\i\k C_{xp}\right)\nonumber \\
 & \hspace{1em}+\i\k\left(x-\mean x\right)\mean p+\i\phi\Bigg),\label{eq:psi_solution}
\end{align}
where $\phi$ is a time dependent global phase. This requires that
the first moments and the variances satisfy the following closed set
of differential equations,
\begin{align}
\Dot{\mean x} & =\mean p,\label{eq:dgl_x}\\
\Dot{\mean p} & =-\mean p,\label{eq:dgl_p}\\
\Dot V_{x} & =C_{xp}-V_{x}\left(4\k^{2}V_{x}-1\right)+\frac{1-\k^{2}C_{xp}^{2}}{16\k^{2}},\label{eq:dgl_v_x}\\
\Dot C_{xp} & =\frac{1+\k^{2}C_{xp}^{2}}{16\k^{2}}\,\frac{8-C_{xp}}{V_{x}}-4\k^{2}C_{xp}V_{x}.\label{eq:dgl_cov}
\end{align}
To see this, one inserts the ansatz (\ref{eq:psi_solution}) into
the NLPSE (\ref{eq:NLPSE QBM}) and compares coefficients in powers
of $x$. A differential equation for $V_{p}$ is readily derived by
noting that the variances of the Gaussian state (\ref{eq:psi_solution})
are related via 
\begin{equation}
4V_{x}V_{p}=\frac{1}{\k^{2}}+C_{xp}^{2}.\label{eq:2mom_connection}
\end{equation}

The equations of motion (\ref{eq:dgl_x}) and (\ref{eq:dgl_p}) for
the first moments exhibit an exponential momentum damping, which reflects
the friction experienced by the Brownian particle through the interaction
with the gas environment. This movement does not affect the form of
the wave packet but only its position in phase space. The equations
of motion (\ref{eq:dgl_v_x}) and (\ref{eq:dgl_cov}) for the variances,
however, drive the width and orientation of the wave packet in phase
space towards the stable fixed values
\begin{align}
V_{x\mathrm{,ps}} & =\frac{1+\sqrt{1+\left(64\k^{2}+1\right)\sqrt{16\k^{2}+1}}}{8\k^{2}\sqrt{16\k^{2}+1}},\label{eq:v_x_fix}\\
C_{xp\mathrm{,ps}} & =\frac{4\k-\sqrt{\sqrt{16\k^{2}+1}-1}}{\k\sqrt{16\k^{2}+1}}.\label{eq:c_xp_fix}
\end{align}
The label ``$\mathrm{ps}$'' denotes the pointer state, and $V_{p\mathrm{,ps}}$
can be calculated with the help of (\ref{eq:2mom_connection}). These
fixed values are obtained by solving the nonlinear equations which
arise by setting $\mathrm{d}V_{x}/\mathrm{d}t=\mathrm{d}C_{xp}/\mathrm{d}t=0$;
their stability can be confirmed by a linear stability analysis of
(\ref{eq:dgl_v_x}) and (\ref{eq:dgl_cov}) around $V_{x\mathrm{,ps}}$
and $C_{xp\mathrm{,ps}}$.

\begin{figure}
\includegraphics[scale=0.3]{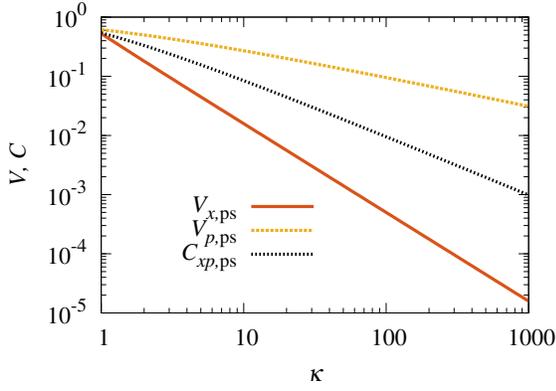}\caption{\label{fig:Scaling-pointers-width}Pointer state variances $V_{x\ind{,ps}}$
in position (solid line) and $V_{p\ind{,ps}}$ momentum space (dashed
line) and covariance $C_{xp,\ind{ps}}$ (dotted line) as a function
of $\k$. For $\k\gg1$, i.e. in the semiclassical limit, all curves
follow a potential law. The scales are defined in Eq.~(\ref{eq:scales}).}
\end{figure}

A wave packet with the fixed values $V_{x,\mathrm{ps}}$ and $V_{p,\mathrm{ps}}$
is a soliton-like solution in the sense that the form of the wave
function in phase space does not change with time. We will call this
solution a \emph{pointer state of quantum Brownian motion} and denote
it by $\ket{\pi\left(\mean x,\mean p\right)}$, where $\mean x$ and
$\mean p$ parametrize the position and momentum coordinates of the
pointer state. As Gaussian wave functions they form an overcomplete
basis set in Hilbert space. From Eqs.~(\ref{eq:2mom_connection})-(\ref{eq:c_xp_fix})
one deduces the asymptotic widths of the pointer states as $\k\rightarrow\infty$,
\begin{equation}
\begin{aligned}V_{x\mathrm{,ps}} & \sim\frac{1}{2\k^{3/2}},\\
V_{p\mathrm{,ps}} & \sim\frac{1}{\k^{1/2}},\\
C_{xp\mathrm{,ps}} & \sim\frac{1}{\k}.
\end{aligned}
\label{eq:scaling of pointer width}
\end{equation}
The pointer states thus get more and more localized in phase space
as one goes deeper into the semiclassical regime ($\hbar\rightarrow0$)
characterized by $\k\rightarrow\infty$, see Eq.~(\ref{eq:k equals}).
This is illustrated in Fig.~\ref{fig:Scaling-pointers-width}.

\subsection{Superposition of wave packets\label{sub:NLPSE superpos o packets}}

We now analyze the action of the NLPSE (\ref{eq:NLPSE QBM}) on a
superposition of separated wave packets. We will see that the NLPSE
suppresses such superpositions until only a single localized wave
packet remains, which eventually turns into a pointer state $\ket{\pi\left(\mean x,\mean p\right)}$
introduced before.

We consider an initial superposition state of the form
\begin{equation}
\ket{\psi}=\sum_{j}c_{j}(0)\ket{\psi_{j}},\label{eq:psi_superpos}
\end{equation}
where the wave packets $\ket{\psi_{j}}$ are well separated in phase
space, orthogonal, and normalized with weights
\begin{equation}
w_{j}(0)=\left|c_{j}(0)\right|^{2},
\end{equation}
summing up to unity. We drop the time argument of the weights and
amplitudes for better readability, and we denote expectation values
of the wave packet components as $\mean A_{j}=\bra{\psi_{j}}\op A\ket{\psi_{j}}$,
and the variances and covariance accordingly.

The $\ket{\psi_{j}}$ are required to be well separated wave packets
in the sense that the orthogonality condition
\begin{equation}
\bra{\psi_{j}}\op A\ket{\psi_{k}}=\delta_{jk}\bra{\psi_{j}}\op A\ket{\psi_{j}}\;\forall j,k,\label{eq:orthogonality}
\end{equation}
holds for the operators $\text{\ensuremath{\op A\in}\{\ensuremath{\hat{\mathbb{I}}}},\op x,\op x^{2},\op p,\op p^{2},\op x\op p+\op p\op x\}$.
The shape of the $\ket{\psi_{j}}$ can be arbitrary, but their widths
in phase space shall be close to that of a pointer state. In particular,
we require that the variances $V_{x,j}$ and $V_{p,j}$ of the wave
packet components $\ket{\psi_{j}}$ exhibit the same dependence (\ref{eq:scaling of pointer width})
on $\k$ as the pointer state. The orthogonality condition (\ref{eq:orthogonality})
can then be fulfilled arbitrarily well by increasing the parameter
$\k$, which decreases the pointer state width in phase space according
to (\ref{eq:scaling of pointer width}), and thus further orthogonalizes
the $\ket{\psi_{j}}$.

For the following, it is useful to express the expectation values
of the superposition state $\ket{\psi}$ with the help of those of
the constituting wave packets $\ket{\psi_{j}}$:
\begin{equation}
\begin{aligned}\mean x & =\sum_{j}w_{j}\mean x_{j},\\
\mean p & =\sum_{j}w_{j}\mean p_{j},\\
V_{x} & =\sum_{j}w_{j}V_{x,j}+\frac{1}{2}\sum_{j,k}w_{j}w_{k}\left(\mean x_{j}-\mean x_{k}\right)^{2},\\
V_{p} & =\sum_{j}w_{j}V_{p,j}+\frac{1}{2}\sum_{j,k}w_{j}w_{k}\left(\mean p_{j}-\mean p_{k}\right)^{2},\\
C_{xp} & =\sum_{j}w_{j}C_{xp,j}+\sum_{j,k}w_{j}w_{k}\left(\mean x_{j}-\mean x_{k}\right)\left(\mean p_{j}-\mean p_{k}\right).
\end{aligned}
\end{equation}
By inserting the superposition state (\ref{eq:psi_superpos}) into
the NLPSE (\ref{eq:NLPSE QBM}) we get the expression
\begin{equation}
\Dot{\ket{\psi}}=\sum_{j}\Dot{\left(c_{j}\right)}\ket{\psi_{j}}+c_{j}\Dot{\ket{\psi_{j}}}=\sum_{j}c_{j}\op N[\psi]\ket{\psi_{j}},
\end{equation}
where the nonlinear operator on the right-hand side contains expectation
values with respect to $\ket{\psi}$. From this equation we deduce
a set of coupled equations for the \emph{normalized} states $\ket{\psi_{j}}$
and their weights $w_{j}=|c_{j}|^{2}$:

\begin{align}
\Dot{\ket{\psi_{j}}} & =\op N[\psi_{j}]\ket{\psi_{j}}+\left(\k\left(\op x-\mean x_{j}\right)+\frac{\i}{4}\left(\op p-\mean p_{j}\right)\right)\nonumber \\
 & \hspace{1em}\times\left(\k\left(\mean x-\mean x_{j}\right)-\frac{\i}{4}\left(\mean p-\mean p_{j}\right)\right)\ket{\psi_{j}},\label{eq:eom psi_j in superpos}\\
\Dot{w_{j}} & =w_{j}\Bigg(2\k^{2}\left(V_{x}-V_{x,j}-\left(\mean x-\mean x_{j}\right)^{2}\right)\nonumber \\
 & \hspace{1em}+\frac{1}{8}\left(V_{p}-V_{p,j}-\left(\mean p-\mean p_{j}\right)^{2}\right)\Bigg).\label{eq:eom w_j in superpos}
\end{align}
Note, that the nonlinear operator appearing in (\ref{eq:eom psi_j in superpos})
now only contains expectation values with respect to $\ket{\psi_{j}}.$
Moreover, if there is only one constituent in the ``superposition'',
Eqs.~(\ref{eq:eom psi_j in superpos}) and (\ref{eq:eom w_j in superpos})
turn consistently into the NLPSE (\ref{eq:NLPSE QBM}) and the trivial
evolution $\mathrm{d}w_{j}/\mathrm{d}t=0$ for the weight. This is
seen most clearly by looking at the differences of the expectation
values, which vanish in that case.

The coupled weight equations (\ref{eq:eom w_j in superpos}) exhibit
a fixed point if a single $w_{j}$ is equal to one, while all others
are zero. It can be checked by a linear stability analysis that this
fixed point is a stable one. Thus, Eq.~(\ref{eq:eom w_j in superpos})
expresses the fact that the NLPSE suppresses superpositions of pointer
states, since the dynamics of the weights always ends up in the stable
fixed point. 

To illustrate the suppression of a superposition state, we consider
a superposition of two pointer states. The weight of the first component
$\ket{\psi_{1}}$ then follows the differential equation
\begin{align}
\Dot{w_{1}} & =\left(2w_{1}-1\right)w_{1}w_{2}\left(2\k^{2}\left(\mean x_{2}-\mean x_{1}\right)^{2}+\frac{1}{8}\left(\mean p_{2}-\mean p_{1}\right)^{2}\right).\label{eq:eom weights two wave packets}
\end{align}
We see immediately that $w_{1}$ gets greater if and only if it is
already the greater of the two weights, leading eventually to the
suppression of the other component. If there exists a superposition
of many wave packets, which, in addition, all have different widths
the situation becomes far more complex and can not be captured intuitively.
However, the eventual decay of the superposition into one of the components
is certain, though it is not easily predictable which component of
the superposition will survive in the course of the evolution due
to Eq.~(\ref{eq:eom w_j in superpos}).

\section{\label{sec:Unraveling-Quantum-Brownian}Unraveling quantum Brownian
motion}

In this section we return to the unraveling of QBM, which consists
of the NLPSE as deterministic part, complemented by stochastic jumps.
This stochastic part is neccessary to produce the required ensemble
of quantum trajetories since the NLPSE alone would always lead to
the same asymptotic pointer state for one particular initial state.
We first show how the jump process leads to stochastically selected
asymptotic pointer states, which occur with relative frequencies according
to Born's rule. After that, in section \ref{sub:Single-Wave-Packets},
we discuss the action of the unraveling on a single pointer state.
The NLPSE describes pointer state trajectories which are exponentially
damped in momentum. We will see how the stochastic part induces a
random walk of the pointer state trajectory displaying the expected
diffusive behavior. In the semiclassical limit this will eventually
yield classical diffusion.

\subsection{\label{sub:Superpos of Wave Packets}Stochastic dynamics of a superposition}

Let us consider the action of a single jump on the superposition state
(\ref{eq:psi_superpos}) of section \ref{sub:NLPSE superpos o packets},
consisting of wave packets $\ket{\psi_{j}}$ that are well separated
in phase space, according to the orthogonality relation (\ref{eq:orthogonality}),
and with widths comparable to the pointer state width,
\begin{equation}
\ket{\psi^{\prime}}=\frac{\op J[\psi]\ket{\psi}}{\sqrt{\langle\op J^{\dagger}\op J\rangle}}=\sum_{j}c_{j}^{\prime}\ket{\psi_{j}^{\prime}}.\label{eq:psi after jump}
\end{equation}
Here, we introduced the normalization factor $\langle\op J^{\dagger}\op J\rangle=\bra{\psi}\op J^{\dagger}[\psi]\op J[\psi]\ket{\psi}$
and the normalized wave packets
\begin{align}
\ket{\psi_{j}^{\prime}} & =\frac{\op J[\psi]\ket{\psi_{j}}}{\sqrt{\langle\op J^{\dagger}\op J\rangle_{j}}},\label{eq:psi_j after jump}
\end{align}
involving the normalization factor 
\begin{align}
\langle\op J^{\dagger}\op J\rangle_{j}=\bra{\psi_{j}}\op J^{\dagger}[\psi]\op J[\psi]\ket{\psi_{j}} & =2\k^{2}\left(V_{x,j}+\left(\mean x-\mean x_{j}\right)^{2}\right)\nonumber \\
 & \hspace{1em}+\frac{1}{8}\left(V_{p,j}+\left(\mean p-\mean p_{j}\right)^{2}\right)-\frac{1}{2}.\label{eq:norm of psi_j after jump}
\end{align}
In the following, only the squared moduli $w_{j}^{\prime}=\left|c_{j}^{\prime}\right|^{2}$
of the new coefficients will be required. One readily finds 
\begin{equation}
w_{j}^{\prime}=\left|c_{j}\right|^{2}\frac{\langle\op J^{\dagger}\op J\rangle_{j}}{\langle\op J^{\dagger}\op J\rangle}.\label{eq:weight after jump}
\end{equation}

Note that the new wave packet components $\ket{\psi_{j}^{\prime}}$
can be safely assumed to be still separated and localized because
the jump operator modifies the shape only linearly in $\op x$ and
$\op p$. The superposition state (\ref{eq:psi after jump}) after
the jump is therefore again a superposition of separated wave packets,
but with different weights. The effect of a jump can thus be approximately
accounted for by a reshuffling of a finite number of weights $w_{j}$
in the superposition. We confirmed this by a numerical implementation
of the unraveling (\ref{eq:ps unraveling QBM}) by combining the Crank-Nicolson
method with the split-operator technique \cite{Fleck1976_FFTMethod,Feit1982_SplitOpFFTSchroedeq,Press2007_NumRec}.

From the action of the NLPSE on a superposition state we calculated
in section \ref{sub:NLPSE superpos o packets} the deterministic evolution
(\ref{eq:eom w_j in superpos}) of the weights $w_{j}$. Knowing the
new weights (\ref{eq:weight after jump}) due to a jump, we are now
in a position to construct a stochastic process for the weights of
the wave packets where the jumps occur with the same rate as in the
original unraveling:

\begin{align}
\mathrm{d}w_{j} & =w_{j}\left(\langle\op J^{\dagger}\op J\rangle-\langle\op J^{\dagger}\op J\rangle_{j}\right)\mathrm{d}t\nonumber \\
 & \hspace{1em}+w_{j}\left(\frac{\langle\op J^{\dagger}\op J\rangle_{j}}{\langle\op J^{\dagger}\op J\rangle}-1\right)\mathrm{d}N.\label{eq:stoch_eq_weights}
\end{align}
Here, we have written the deterministic part (\ref{eq:eom w_j in superpos})
more conveniently with the help of the jump rate $r=\langle\op J^{\dagger}\op J\rangle$
(\ref{eq:jump rate QBM}) and the normalization factor $\langle\op J^{\dagger}\op J\rangle_{j}$
(\ref{eq:norm of psi_j after jump}). The Poisson increment $\mathrm{d}N$
has the same ensemble average $\mathrm{E}[\mathrm{d}N]=r\,\mathrm{d}t$
as the pointer state unraveling (\ref{eq:ps unraveling QBM}).

The set of equations (\ref{eq:stoch_eq_weights}) represent the evolution
of the weights in a quantum trajectory. Each quantum trajectory ends
up asymptotically in a pointer state, when all weights but a single
one vanish. From this we deduce that the relative frequency of finding
a particular pointer state $\ket{\pi\left(\mean x_{j},\mean p_{j}\right)}$
is equal to the probability of the associated asymptotic state in
the ensemble of quantum trajectories. This latter probability is evaluated
easily by carrying out the ensemble average of (\ref{eq:stoch_eq_weights}),
\begin{align}
\mathrm{E}\left[\mathrm{d}w_{j}\right] & =0.
\end{align}
This implies $\mathrm{E}\left[w_{j}(t)\right]=\mathrm{const}.$ and,
in particular,
\begin{equation}
\mathrm{E}\left[w_{j}(t\rightarrow\infty)\right]=\mathrm{E}\left[w_{j}(0)\right]=|c_{j}(0)|^{2},\label{eq:dgl_average_weights}
\end{equation}
where $c_{j}(0)$ are the initial amplitudes of the wave packets in
the superposition state (\ref{eq:psi_superpos}). Equation (\ref{eq:dgl_average_weights})
confirms Born's rule (\ref{eq:prob_for_ps}), as expected for a superposition
of separated wave packets.

\subsection{\label{sub:Single-Wave-Packets}Stochastic dynamics of a single wave
packet}

So far, we investigated the dynamics of the unraveling if the initial
state is a superposition of localized wave packets. We confirmed in
the previous section that the probability to end up asymptotically
in a certain pointer state is given by its weight in the initial superposition
(Born's rule). This process takes place on the fast decoherence time
scale $t_{\mathrm{dec}}$ after which the system is well described
by a pointer state. In terms of the pointer state unraveling, this
point is reached when each quantum trajectory in the ensemble has
turned into its asymptotic pointer state.

Since the underlying quantum dynamics is diffusive \cite{Breuer_Petruccione_OpenQS}
it is natural to describe the statistics of the position $\mean x$
and the momentum $\mean p$ of the pointer state by the phase space
diffusion
\begin{equation}
\left(\begin{aligned}\mathrm{d}\mean x\\
\mathrm{d}\mean p
\end{aligned}
\right)=\left(\begin{aligned}\mean p\\
-\mean p
\end{aligned}
\right)\mathrm{d}t+\left(\begin{array}{cc}
B_{11} & B_{12}\\
B_{21} & B_{22}
\end{array}\right)\left(\begin{array}{c}
\mathrm{d}W_{1}\\
\mathrm{d}W_{2}
\end{array}\right),\label{eq:class phase space diffusion}
\end{equation}
with the Wiener increments $\mathrm{d}W_{1}$ and $\mathrm{d}W_{2}$
obeying the rules 
\begin{equation}
\begin{aligned}\mathrm{E}[\mathrm{d}W_{i}] & =0,\\
\mathrm{d}W_{i}\mathrm{d}W_{j} & =\delta_{ij}\mathrm{d}t.
\end{aligned}
\label{eq: Wiener inc rule}
\end{equation}
Equation (\ref{eq:class phase space diffusion}) is a special case
of the two-dimensional Ornstein-Uhlenbeck process with constant coefficients
\cite{gardiner2004handbook} and the derivation of its ensemble variances
and covariance is presented in the appendix. The different diffusion
constants $D_{x}$, $D_{p}$, and $D_{xp}$ describing position, momentum,
and covariance diffusion are related to the $B_{ij}$ by
\begin{equation}
\begin{aligned}D_{x} & =B_{11}^{2}+B_{12}^{2},\\
D_{p} & =B_{21}^{2}+B_{22}^{2},\\
D_{xp} & =B_{11}B_{21}+B_{12}B_{22}.
\end{aligned}
\label{eq: diff consts via wiener proc}
\end{equation}

We proceed to show how the stochastic jumps acting on the motion of
the pointer state lead to phase space trajectories which turn into
classical diffusion in the semiclassical limit $\k\rightarrow\infty$.
To achieve this, we derive a simple model allowing the description
of the pointer state trajectories by the phase space diffusion (\ref{eq:class phase space diffusion})
and then compare it to a numerical study.

\subsubsection{Analytic model}

\begin{figure*}
\includegraphics[scale=0.4]{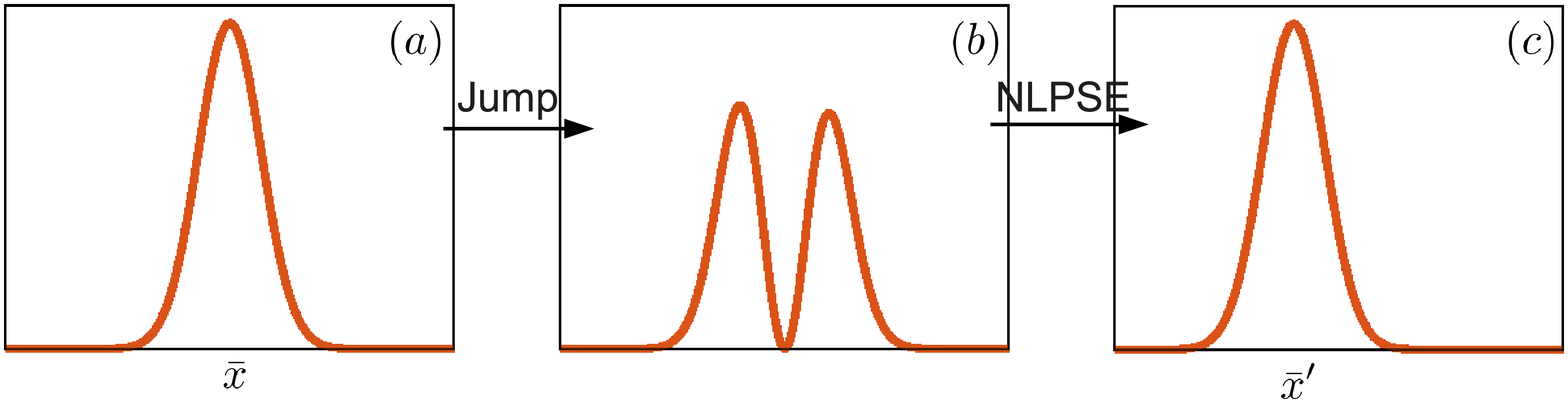}

\caption{\label{fig:3stepjumpaction}Simplified jump process: A jump turns
the wave packet at position $\mean x$ $(a)$ into a slightly asymmetric
double-peaked structure $(b)$. The NLPSE then suppresses one of the
peaks resulting in a single wave packet at the position $\mean x^{\prime}$
of one of the double-peaks $(c)$. A similar picture applies in momentum
space.}
\end{figure*}

Consider a pointer state moving on the phase space trajectory governed
by the NLPSE (see Sec.~\ref{sub:NLPSE Single Wave Packets}). This
deterministic motion is interrupted by a jump, described by the jump
operator (\ref{eq:jump op QBM}) acting on the pointer state (\ref{eq:psi_solution}),
\begin{align}
\psi^{\prime}(x) & =\bra x\op J[\pi\left(\mean x,\mean p\right)]\ket{\pi\left(\mean x,\mean p\right)}\nonumber \\
 & =\sqrt{\frac{2}{r_{\ind{ps}}}}\left[\k\left(1-\frac{1-\i\k C_{xp\mathrm{,ps}}}{8\k^{2}V_{x\mathrm{,ps}}}\right)\left(x-\mean x\right)\right]\langle x\ket{\pi\left(\mean x,\mean p\right)}.\label{eq:jumpaction on ps}
\end{align}
Here, 
\begin{equation}
r_{\mathrm{ps}}=2\k^{2}V_{x,\ind{ps}}+\frac{1}{8}V_{p,\ind{ps}}-\frac{1}{2}\label{eq: jump rate ps}
\end{equation}
is the jump rate (\ref{eq:jump rate QBM}) of the pointer state and
$V_{x,\ind{ps}}$ and $V_{p,\ind{ps}}$ are the widths of the pointer
states defined in Eqs.~(\ref{eq:2mom_connection})-(\ref{eq:c_xp_fix}).
The resulting state is a \emph{symmetric} double-peaked wave packet.
This symmetry reflects the assumed symmetry of the state before the
jump. Since in reality the system state is never exactly symmetric
even small asymmetries will trigger the emergence of a single wave
packet under the NLPSE, see Fig.~\ref{fig:3stepjumpaction}.

Which of the two subpeaks survives depends on the asymmetry before
the jump. We assume equal probabilities and that the single wave packet
is restored sufficiently fast by the NLPSE. This results in an effective
jump of the pointer state's first moments in phase space by the distances
\begin{equation}
\begin{aligned}j_{x} & =|\mean x^{\prime}-\mean x|=\sqrt{2V_{x\mathrm{,ps}}},\\
j_{p} & =|\mean p^{\prime}-\mean p|=\sqrt{2V_{p\mathrm{,ps}}}.
\end{aligned}
\label{eq: jump width analytic}
\end{equation}
The widths $j_{x}$ and $j_{p}$ are calculated from the positions
of the two peaks in (\ref{eq:jumpaction on ps}). Since the direction
of the jump is positive or negative with equal probability, we end
up with the two possible jumps in phase space
\begin{equation}
\begin{aligned}\vec{j}_{1} & =\left(\begin{array}{c}
j_{x}\\
j_{p}
\end{array}\right), & \vec{j}_{2} & =\left(\begin{array}{c}
-j_{x}\\
-j_{p}
\end{array}\right),\end{aligned}
\label{eq:classical jumps}
\end{equation}
each of them characterized by a Poisson process $\mathrm{d}N_{k}$
with jump rate
\begin{align}
r_{k} & =\frac{\mathrm{E}[\mathrm{d}N_{k}]}{\mathrm{d}t}=\frac{r_{\mathrm{ps}}}{2}.\label{eq:classical rate}
\end{align}

We are now in the position to write down a stochastic differential
equation for $\mean x$ and $\mean p$,
\begin{align}
\left(\begin{aligned}\mathrm{d}\mean x\\
\mathrm{d}\mean p
\end{aligned}
\right) & =\left(\begin{aligned}\mean p\\
-\mean p
\end{aligned}
\right)\mathrm{d}t+\sum_{k=1}^{2}\vec{j}_{k}\mathrm{d}N_{k}.\label{eq:analytic poisson SDE}
\end{align}
It consists of the deterministic evolution (\ref{eq:dgl_x}) and (\ref{eq:dgl_p})
derived in section \ref{sub:NLPSE Single Wave Packets} and the stochastic
jumps (\ref{eq:classical jumps}) with corresponding jump rates (\ref{eq:classical rate}).
By using the Poisson rule $\mathrm{d}N_{k}\mathrm{d}N_{l}=\delta_{kl}\mathrm{d}N_{k}$
and Eq.~(\ref{eq:classical rate}), one can easily derive all moments
of the stochastic process (\ref{eq:analytic poisson SDE}),
\begin{align}
 & \mathrm{E}\left[\left(\mathrm{d}x\right)^{2n-1}\right]=0, & \mathrm{E}\left[\left(\mathrm{d}x\right)^{2n}\right]=j_{x}^{2n}r_{\ind{ps}}\mathrm{d}t,\nonumber \\
 & \mathrm{E}\left[\left(\mathrm{d}p\right)^{2n-1}\right]=0, & \mathrm{E}\left[\left(\mathrm{d}p\right)^{2n}\right]=j_{p}^{2n}r_{\ind{ps}}\mathrm{d}t,\label{eq: moments of stoch process}\\
 & \mathrm{E}\left[\left(\mathrm{d}x\mathrm{d}p\right)^{n}\right]=\left(j_{x}j_{p}\right)^{n}r_{\ind{ps}}\mathrm{d}t,\nonumber 
\end{align}
with $n\ge1$. If one is interested in calculating moments up to second
order only, the stochastic jump process (\ref{eq:analytic poisson SDE})
can be well approximated by the phase space diffusion (\ref{eq:class phase space diffusion})
upon identifying the diffusion constants (\ref{eq: diff consts via wiener proc})
with
\begin{align}
D_{x} & =j_{x}^{2}r_{\ind{ps}},\nonumber \\
D_{p} & =j_{p}^{2}r_{\ind{ps}},\label{eq: diff consts via jump proc}\\
D_{xp} & =j_{x}j_{p}r_{\ind{ps}}.\nonumber 
\end{align}

Inserting the jump widths (\ref{eq: jump width analytic}) and the
jump rate (\ref{eq: jump rate ps}) into Eqs.~(\ref{eq: diff consts via jump proc})
one gets their dependence on $\k$ as $\k\rightarrow\infty$
\begin{equation}
\begin{aligned}D_{x} & \sim\frac{1}{\k},\\
D_{p} & \sim2,\\
D_{xp} & \sim\frac{1}{\k^{1/2}}.
\end{aligned}
\label{eq:scaling of diff consts}
\end{equation}
We see that both $D_{x}$ and $D_{xp}$ vanish in the semiclassical
limit $\k\rightarrow\infty$ so that only the momentum diffusion $D_{p}=2$
contributes and the stochastic process (\ref{eq:class phase space diffusion})
thus reduces to

\begin{align}
\left(\begin{aligned}\mathrm{d}\mean x\\
\mathrm{d}\mean p
\end{aligned}
\right) & =\left(\begin{aligned}\mean p\\
-\mean p
\end{aligned}
\right)\mathrm{d}t+\left(\begin{aligned}0\\
\sqrt{D_{p}}
\end{aligned}
\right)\mathrm{d}W.\label{eq:Langevin equation CBM}
\end{align}
One can also arrive directly at (\ref{eq:Langevin equation CBM})
without defining the phase space diffusion (\ref{eq:class phase space diffusion})
by drawing the semiclassical limit of Eq.~(\ref{eq:analytic poisson SDE})
and thereby performing a diffusive limit analogous to that of a random
walker \cite{gardiner2004handbook}. Writing (\ref{eq:Langevin equation CBM})
in physical dimensions by using the scales defined in Eq.~(\ref{eq:scales})
finally leads to the Langevin equation of classical Brownian motion
\cite{Risken1989_FPE,Breuer_Petruccione_OpenQS}
\begin{equation}
\mathrm{d}\mean p=-2\gamma\mean p\mathrm{d}t+\sqrt{4\gamma mk_{\ind B}T_{\ind{env}}}\mathrm{d}W,\label{eq: Langevin eq w dimension}
\end{equation}
where the dimension of $\mathrm{d}W$ is a square root of time.

It should be remarked here, that the crucial assumption leading to
this model is the \emph{immediate} restoring of the pointer state
after a jump. The real dynamics will take a finite time until the
double-peaked wave function returns to a pointer state, and there
will be several jumps in between. We show in the following, by numerically
simulating trajectories of the pointer state unraveling, that this
is indeed the case, giving rise to complex dynamics. Nonetheless,
the general picture of the analytic model can be confirmed.

\subsubsection{Numerical study}

Using a combination of the Crank-Nicolson method and a split-operator
technique \cite{Fleck1976_FFTMethod,Feit1982_SplitOpFFTSchroedeq,Press2007_NumRec}
we generated quantum trajectories of the unraveling (\ref{eq:ps unraveling QBM})
and calculated their first moments $\mean x$ and $\mean p$ and their
variances $V_{x}$, $V_{p}$ and $C_{xp}$. 
\begin{figure*}
\includegraphics[scale=0.65]{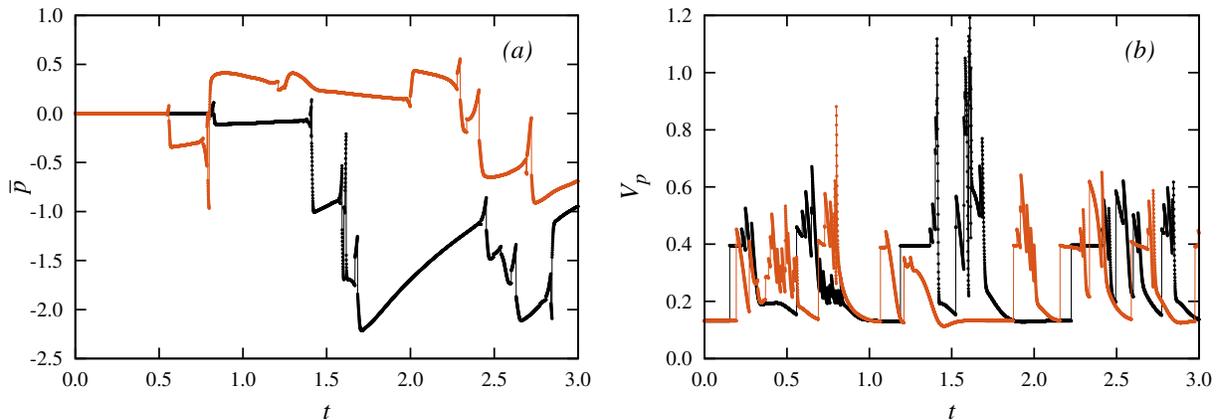}

\caption{\label{fig:2 sample traj}$(a)$ momentum expectation $\mean p$ and
$(b)$ momentum variance $V_{p}$ of two sample trajectories starting
at the same pointer state $\ket{\pi\left(0,0\right)}$. One observes
that in general more than a single jump occurs in the unraveling before
the width of the wave function is restored to the pointer state width.
The calculations are made at $\k=50$ and all quantities are dimensionless
with scales defined in Eq.~(\ref{eq:scales}).}
\end{figure*}
Figure \ref{fig:2 sample traj} shows exemplarily the temporal evolution
of $\mean p$ and $V_{p}$ for two sample trajectories. A similar
behavior is found in position space (not shown).

\paragraph{Verification of the stochastic model}

In order to compare the statistics of the phase space trajectories
predicted by the phase space diffusion (\ref{eq:class phase space diffusion})
with that of the numerically generated trajectories we consider the
expectation values of a finite sample of size $N$
\begin{align}
\mathrm{E}_{N}[x] & =\frac{1}{N}\sum_{i=1}^{N}\mean x_{i},\nonumber \\
\mathrm{E}_{N}[p] & =\frac{1}{N}\sum_{i=1}^{N}\mean p_{i},\nonumber \\
\mathrm{Var}_{N}[x] & =\frac{1}{N-1}\sum_{i=1}^{N}\left(\mean x_{i}-\mathrm{E}_{N}[x]\right)^{2},\label{eq:sample means}\\
\mathrm{Var}_{N}[p] & =\frac{1}{N-1}\sum_{i=1}^{N}\left(\mean p_{i}-\mathrm{E}_{N}[p]\right)^{2},\nonumber \\
\mathrm{Cov}_{N}[x,p] & =\frac{1}{N-1}\sum_{i=1}^{N}\left(\mean x_{i}-\mathrm{E}_{N}[x]\right)\left(\mean p_{i}-\mathrm{E}_{N}[p]\right),\nonumber 
\end{align}
where $\mean x_{i}$, $\mean p_{i}$ are the first moments of quantum
trajectory $i$. The sample expectation values (\ref{eq:sample means})
are themselves stochastic values due to the stochasticity of the underlying
model. The ensemble variances of $\mathrm{E}_{N}\left[x\right]$ and
$\mathrm{E}_{N}\left[p\right]$ on the one hand and those of $\mathrm{Var}_{N}[x]$,
$\mathrm{Var}_{N}[p]$, and $\mathrm{Cov}_{N}[x,p]$ on the other
are proportional to $1/N$ and $1/\left(N-1\right)$, respectively
(see appendix). We compare the finite-size variances of the numerically
generated trajectories with those predicted by the above model.
\begin{figure*}
\includegraphics[scale=0.65]{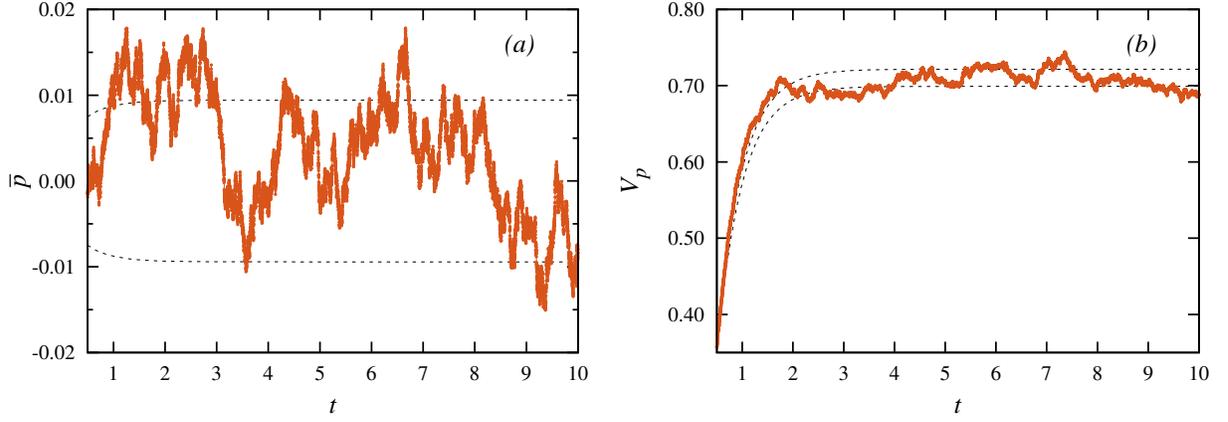}

\caption{\label{fig:sample means p}$\mathrm{E}_{N}\left[p\right]$ and $\mathrm{Var}_{N}\left[p\right]$
(solid) as well as their theoretically expected deviations (dashed)
due to a finite sample size. The fluctuations are well characterized
by the dashed lines. Calculations are done at $\k=50$ for $N=8000$
trajectories.}
\end{figure*}
 Figure \ref{fig:sample means p} compares $\mathrm{E}_{N}\left[p\right]$
and $\mathrm{Var}_{N}\left[p\right]$ of the sample with the theoretically
expected deviations. The interval of one standard deviation around
the ensemble mean value $\mathrm{E}_{\infty}\left[p\right]$ should
contain $\approx68\%$ of the trajectories of the sample, which we
confirmed at selected times. The same analysis applies for $\mathrm{E}_{N}[x]$,
$\mathrm{Var}_{N}[x]$, and $\mathrm{Cov}_{N}[x,p]$ and yields the
same result. Thus, we conclude that the statistics of the moments
of the pointer state unraveling can indeed be described by the stochastic
model (\ref{eq:class phase space diffusion}).

\paragraph{Extraction of the diffusion constants}

\begin{figure}
\includegraphics[scale=0.3]{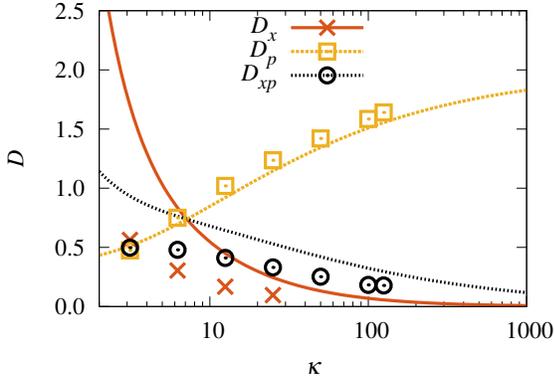}

\caption{\label{fig:Scaling Diff with kappa}Dependence of the diffusion constants
$D_{x}$ (solid line), $D_{p}$ (dashed line) and $D_{xp}$ (dotted
line) on $\k$. The lines are calculated from Eqs.~(\ref{eq: diff consts via jump proc})
of the analytic model, whereas the symbols represent the results of
the numerical simulation. One observes for all three diffusion constants
that the analytic model captures the main features of their dependence
on $\k$. Specifically, $D_{p}$ approaches the classical value of
2 in the semiclassical regime $\k\rightarrow\infty$, whereas $D_{x}$
and $D_{xp}$ vanish in that limit.}
\end{figure}

Having verified the stochastic model we are now able to extract the
diffusion constants $D_{x}$, $D_{p}$ and $D_{xp}$ by fitting the
variances shown in Eqs.~(\ref{eq:A Solution 2nd moments}) to the
generated trajectories. Since $\mathrm{Var}[p]$ only depends on $D_{p}$,
and $\mathrm{Cov}[x,p]$ only on $D_{p}$ and $D_{xp}$, fitting is
done consecutively by extracting $D_{p}$ from $\mathrm{Var}[p]$,
$D_{xp}$ from $\mathrm{Cov}[x,p]$, and finally $D_{x}$ from $\mathrm{Var}[x]$.
The numerically obtained values for different $\k$, as well as the
calculations from the analytic model (\ref{eq:scaling of diff consts}),
are shown in Fig.~\ref{fig:Scaling Diff with kappa}.
 We see that the calculated $D_{p}$ is in good quantitative agreement
with the simulation. The deviation of the calculated $D_{x}$ and
$D_{xp}$ from the simulated ones is due to the crude simplification
made by the assumption that a jump in the unraveling corresponds to
a phase-space jump of the pointer state without further dynamics;
Fig.~\ref{fig:2 sample traj} shows clearly that this is not the
case.

For large $\k$ both the analytic model and the numerical simulation
exhibit the classically expected behavior. Specifically, the position
diffusion $D_{x}$ and the covariance diffusion $D_{xp}$ tend to
zero, whereas the momentum diffusion $D_{p}$ approaches the value
2.

\section{\label{sec:Conclusion}Conclusion}

We identified the pointer states of quantum Brownian motion and derived
the stochastic equations of motion for their trajectories in phase
space. The pointer states turn out to be Gaussian wave packets with
fixed widths and a finite position-momentum covariance. The phase
space trajectory characterizing their motion obeys a stochastic differential
equation combining momentum damping with position and momentum diffusion.
In the semiclassical limit, it turns into the Langevin equation of
classical diffusion. This provides us with a consistent and transparent
picture of the quantum-classical transition as induced by quantum
Brownian motion.

We saw that any initial state, which can be represented by a superposition
of sufficiently separated wave packets decays into a mixture of corresponding
pointer states, and that the ensuing statistical weights are consistent
with Born's rule. Using the pointer states and their trajectories
one can thus capture an essential part of the dynamics of quantum
Brownian motion. Our analysis depends crucially on linking the nonlinear
equation of motion for the pointer states to a particular piecewise
deterministic unraveling of the master equation. It allows one to
connect the open quantum dynamics to stochastic trajectories in phase
space.

The described technique of unraveling a master equation into pointer
state trajectories may become particularly useful in cases where even
approximate solutions of a master equation are not known. For example,
the full Markovian dynamics of a tracer particle in a gaseous environment
is described by a complicated integro-differential equation not readily
accessible by analytic means \cite{Vacchini_Hornberger_QLBE_Review}.
If pointer states can be identified for this equation we expect their
trajectories to be described by a general Chapman-Kolmogorov equation
in phase space turning into the linear Boltzmann equation in the semiclassical
limit.

This work was supported by the DFG via SFB/TR 12.

\appendix

\section{Diffusion in phase space\label{sec:Appendix Diffusion-in-Phase}}

We analyze the properties of the classical stochastic process
\begin{equation}
\begin{aligned}\mathrm{d}x & =p\,\mathrm{d}t+B_{11}\mathrm{d}W_{1}+B_{12}\mathrm{d}W_{2},\\
\mathrm{d}p & =-p\,\mathrm{d}t+B_{21}\mathrm{d}W_{1}+B_{22}\mathrm{d}W_{2},
\end{aligned}
\label{eq:A classical phase space diffusion}
\end{equation}
used in section \ref{sub:Single-Wave-Packets} to describe the dynamics
of the first moments $\mean x$, $\mean p$ of the pointer states
in phase space. At first, we discuss the ensemble statistics, and
afterwards the statistics of a sample of finite size.

\subsection{Ensemble statistics}

In order to derive the statistical properties of the stochastic process
(\ref{eq:A classical phase space diffusion}), we recall the stochastic
properties of the real-valued Wiener increments $\mathrm{d}W_{1}$
and $\mathrm{d}W_{2}$
\begin{equation}
\begin{aligned}\mathrm{E}[\mathrm{d}W_{i}] & =0,\\
\mathrm{d}W_{i}\mathrm{d}W_{j} & =\delta_{ij}\mathrm{d}t.
\end{aligned}
\label{eq:A properties Wiener increments}
\end{equation}
From these two relations one can derive a differential equation for
e.g. $x^{2}$ by expanding to second order in $\mathrm{d}x$ and keeping
only terms to the order $\mathrm{d}t$, known as the Ito rule of stochastic
calculus. One gets
\begin{equation}
\begin{aligned}\mathrm{d}\left(x^{2}\right) & =2x\mathrm{d}x+\left(\mathrm{d}x\right)^{2}\\
 & =\left(2xp+B_{11}^{2}+B_{12}^{2}\right)\mathrm{d}t+2x\left(B_{11}\mathrm{d}W_{1}+B_{12}\mathrm{d}W_{2}\right).
\end{aligned}
\end{equation}
In the same manner one calculates $\mathrm{d}\left(p^{2}\right)$
and $\mathrm{d}\left(xp\right)$, which then allows the derivation
of differential equations for the ensemble variances $\mathrm{Var}[x]=\mathrm{E}[x^{2}]-\mathrm{E}[x]^{2}$
and $\mathrm{Var}[p]=\mathrm{E}[p^{2}]-\mathrm{E}[p]^{2}$ and the
ensemble covariance $\mathrm{Cov}[x,p]=\mathrm{E}[xp]-\mathrm{E}[x]\mathrm{E}[p]$:
\begin{equation}
\begin{aligned}\frac{\mathrm{d}}{\mathrm{d}t}\mathrm{Var}[x] & =2\mathrm{Cov}[x,p]+D_{x},\\
\frac{\mathrm{d}}{\mathrm{d}t}\mathrm{Var}[p] & =-2\mathrm{Var}[p]+D_{p},\\
\frac{\mathrm{d}}{\mathrm{d}t}\mathrm{Cov}[x,p] & =-\mathrm{Cov}[x,p]+\mathrm{Var}[p]+D_{xp},
\end{aligned}
\label{eq:A DGLs 2nd moments}
\end{equation}
where definition (\ref{eq: diff consts via wiener proc}) of the diffusion
constants was used. Solving the differential equations (\ref{eq:A DGLs 2nd moments})
for initial conditions $\mathrm{Var}[x](0)=\mathrm{Var}[p](0)=\mathrm{Cov}[x,p](0)=0$
yields
\begin{equation}
\begin{aligned}\mathrm{Var}[x](t) & =\left(D_{x}+D_{p}+2D_{xp}\right)t-\frac{1}{2}D_{p}\left(1-\mathrm{e}^{-t}\right)^{2}\\
 & \hspace{1em}-\left(D_{p}+2D_{xp}\right)\left(1-\mathrm{e}^{-t}\right),\\
\mathrm{Var}[p](t) & =\frac{1}{2}D_{p}\left(1-\mathrm{e}^{-2t}\right),\\
\mathrm{Cov}[x,p](t) & =\frac{1}{2}D_{p}\left(1-\mathrm{e}^{-t}\right)^{2}+D_{xp}\left(1-\mathrm{e}^{-t}\right).
\end{aligned}
\label{eq:A Solution 2nd moments}
\end{equation}

\subsection{Statistics with finite sample size}

We are now interested in the statistics of the first moments and the
variances of a finite sample of size $N$, and, in particular, their
deviations around the ensemble values (\ref{eq:A Solution 2nd moments}).
With the help of the stochastic model (\ref{eq:A classical phase space diffusion})
and the sample expectation values (\ref{eq:sample means}) one can
write down e.g. the stochastic differential equation for the sample
momentum expectation
\begin{align}
\mathrm{d}\mathrm{E}_{N}[p] & =-\mathrm{E}_{N}[p]\mathrm{d}t+\frac{1}{N}\sum_{i=1}^{N}\left(B_{21}\mathrm{d}W_{1,i}+B_{22}\mathrm{d}W_{2,i}\right),\label{eq:A SDGL E_N p}
\end{align}
with independent Wiener increments $\mathrm{d}W_{1,i}$ and $\mathrm{d}W_{2,i}$
for all $i$. In the same manner as in the previous section one can
then derive
\begin{equation}
\begin{aligned}\mathrm{d}\left(\mathrm{E}_{N}[p]^{2}\right) & =2\mathrm{E}_{N}[p]\mathrm{d}\mathrm{E}_{N}[p]+\left(\mathrm{d}\mathrm{E}_{N}[p]\right)^{2}\\
 & =\left(-2\mathrm{E}_{N}[p]^{2}+\frac{B_{21}^{2}+B_{22}^{2}}{N}\right)\mathrm{d}t\\
 & \hspace{1em}+\frac{2}{N}\mathrm{E}_{N}[p]\sum_{i=1}^{N}\left(B_{21}\mathrm{d}W_{1,i}+B_{22}\mathrm{d}W_{2,i}\right),
\end{aligned}
\end{equation}
and finally

\begin{align}
\frac{\mathrm{d}}{\mathrm{d}t}\mathrm{Var}[\mathrm{E}_{N}[p]] & =-2\mathrm{Var}[\mathrm{E}_{N}[p]]+\frac{D_{p}}{N}.\label{eq:A DGL Var E_N p}
\end{align}
Here, the ensemble variance of $\mathrm{E}_{N}\left[p\right]$ is
denoted by $\mathrm{Var}[\mathrm{E}_{N}[p]]=\mathrm{E}[\mathrm{E}_{N}[p]^{2}]-\mathrm{E}[\mathrm{E}_{N}[p]]^{2}$
and the definition of the diffusion constants (\ref{eq: diff consts via wiener proc})
was used.

Similarly, one derives a stochastic differential equation for the\emph{
}sample\emph{ }variance of the momentum
\begin{align}
\mathrm{d}\mathrm{Var}_{N}[p] & =\left(D_{p}-2\mathrm{Var}_{N}[p]\right)\mathrm{d}t+\frac{2}{N-1}\sum_{i=1}^{N}\left(p_{i}-\mathrm{E}_{N}[p]\right)\nonumber \\
 & \hspace{1em}\times\left(B_{21}\mathrm{d}W_{1,i}+B_{22}\mathrm{d}W_{2,i}\right),\label{eq:A SDGL Var_N p}
\end{align}
yielding 
\begin{align}
\frac{\mathrm{d}}{\mathrm{d}t}\mathrm{Var}[\mathrm{Var}_{N}[p]] & =-4\mathrm{Var}[\mathrm{Var}_{N}[p]]+\frac{4D_{p}\mathrm{Var}[p]}{N-1}
\end{align}
for the \emph{ensemble} variance of $\mathrm{Var}_{N}[p]$. The variances
of $\mathrm{E}_{N}[x]$, $\mathrm{Var}_{N}[x]$ and $\mathrm{Cov}_{N}[x,p]$
are obtained accordingly in a long and tedious calculation, giving
the solutions
\begin{equation}
\begin{aligned}\mathrm{Var}[\mathrm{E}_{N}[x]] & =\frac{1}{N}\mathrm{Var}[x],\\
\mathrm{Var}[\mathrm{E}_{N}[p]] & =\frac{1}{N}\mathrm{Var}[p],\\
\mathrm{Var}[\mathrm{Var}_{N}[x]] & =\frac{2}{N-1}\mathrm{Var}[x]^{2},\\
\mathrm{Var}[\mathrm{Var}_{N}[p]] & =\frac{2}{N-1}\mathrm{Var}[p]^{2},\\
\mathrm{Var}[\mathrm{Cov}_{N}[x,p]] & =\frac{1}{N-1}\left(\mathrm{Var}[x]\mathrm{Var}[p]+\mathrm{Cov}[x,p]^{2}\right).
\end{aligned}
\end{equation}


\begin{thebibliography}{42}%
\makeatletter
\providecommand \@ifxundefined [1]{%
 \@ifx{#1\undefined}
}%
\providecommand \@ifnum [1]{%
 \ifnum #1\expandafter \@firstoftwo
 \else \expandafter \@secondoftwo
 \fi
}%
\providecommand \@ifx [1]{%
 \ifx #1\expandafter \@firstoftwo
 \else \expandafter \@secondoftwo
 \fi
}%
\providecommand \natexlab [1]{#1}%
\providecommand \enquote  [1]{``#1''}%
\providecommand \bibnamefont  [1]{#1}%
\providecommand \bibfnamefont [1]{#1}%
\providecommand \citenamefont [1]{#1}%
\providecommand \href@noop [0]{\@secondoftwo}%
\providecommand \href [0]{\begingroup \@sanitize@url \@href}%
\providecommand \@href[1]{\@@startlink{#1}\@@href}%
\providecommand \@@href[1]{\endgroup#1\@@endlink}%
\providecommand \@sanitize@url [0]{\catcode `\\12\catcode `\$12\catcode
  `\&12\catcode `\#12\catcode `\^12\catcode `\_12\catcode `\%12\relax}%
\providecommand \@@startlink[1]{}%
\providecommand \@@endlink[0]{}%
\providecommand \url  [0]{\begingroup\@sanitize@url \@url }%
\providecommand \@url [1]{\endgroup\@href {#1}{\urlprefix }}%
\providecommand \urlprefix  [0]{URL }%
\providecommand \Eprint [0]{\href }%
\providecommand \doibase [0]{http://dx.doi.org/}%
\providecommand \selectlanguage [0]{\@gobble}%
\providecommand \bibinfo  [0]{\@secondoftwo}%
\providecommand \bibfield  [0]{\@secondoftwo}%
\providecommand \translation [1]{[#1]}%
\providecommand \BibitemOpen [0]{}%
\providecommand \bibitemStop [0]{}%
\providecommand \bibitemNoStop [0]{.\EOS\space}%
\providecommand \EOS [0]{\spacefactor3000\relax}%
\providecommand \BibitemShut  [1]{\csname bibitem#1\endcsname}%
\let\auto@bib@innerbib\@empty
\bibitem [{\citenamefont {Joos}\ \emph {et~al.}(2003)\citenamefont {Joos},
  \citenamefont {Zeh}, \citenamefont {Kiefer}, \citenamefont {Giulini},
  \citenamefont {Kupsch},\ and\ \citenamefont
  {Stamatescu}}]{Joos_Deco_App_Class_World}%
  \BibitemOpen
  \bibfield  {author} {\bibinfo {author} {\bibfnamefont {E.}~\bibnamefont
  {Joos}}, \bibinfo {author} {\bibfnamefont {H.~D.}\ \bibnamefont {Zeh}},
  \bibinfo {author} {\bibfnamefont {C.}~\bibnamefont {Kiefer}}, \bibinfo
  {author} {\bibfnamefont {D.}~\bibnamefont {Giulini}}, \bibinfo {author}
  {\bibfnamefont {J.}~\bibnamefont {Kupsch}}, \ and\ \bibinfo {author}
  {\bibfnamefont {I.~O.}\ \bibnamefont {Stamatescu}},\ }\href@noop {} {\emph
  {\bibinfo {title} {{Decoherence and the Appearance of a Classical World in
  Quantum Theory}}}}\ (\bibinfo  {publisher} {Springer-Verlag Berlin},\
  \bibinfo {year} {2003})\BibitemShut {NoStop}%
\bibitem [{\citenamefont
  {Zurek}(2003)}]{Zurek_Review_Dec_Eins_Quant_Origin_Class}%
  \BibitemOpen
  \bibfield  {author} {\bibinfo {author} {\bibfnamefont {W.~H.}\ \bibnamefont
  {Zurek}},\ }\href {\doibase 10.1103/RevModPhys.75.715} {\bibfield  {journal}
  {\bibinfo  {journal} {Rev. Mod. Phys.}\ }\textbf {\bibinfo {volume} {75}},\
  \bibinfo {pages} {715} (\bibinfo {year} {2003})}\BibitemShut {NoStop}%
\bibitem [{\citenamefont {Schlosshauer}(2007)}]{MSchlosshBook}%
  \BibitemOpen
  \bibfield  {author} {\bibinfo {author} {\bibfnamefont {M.}~\bibnamefont
  {Schlosshauer}},\ }\href@noop {} {\emph {\bibinfo {title} {{Decoherence and
  the quantum to classical transition}}}}\ (\bibinfo  {publisher}
  {Springer-Verlag Berlin},\ \bibinfo {year} {2007})\BibitemShut {NoStop}%
\bibitem [{\citenamefont {Zurek}(1981)}]{Zurek_What_Mixtures_PS}%
  \BibitemOpen
  \bibfield  {author} {\bibinfo {author} {\bibfnamefont {W.~H.}\ \bibnamefont
  {Zurek}},\ }\href {\doibase 10.1103/PhysRevD.24.1516} {\bibfield  {journal}
  {\bibinfo  {journal} {Phys. Rev. D}\ }\textbf {\bibinfo {volume} {24}},\
  \bibinfo {pages} {1516} (\bibinfo {year} {1981})}\BibitemShut {NoStop}%
\bibitem [{\citenamefont {Zurek}\ \emph {et~al.}(1993)\citenamefont {Zurek},
  \citenamefont {Habib},\ and\ \citenamefont
  {Paz}}]{Zurek_CohStatesViaDecoh1993}%
  \BibitemOpen
  \bibfield  {author} {\bibinfo {author} {\bibfnamefont {W.~H.}\ \bibnamefont
  {Zurek}}, \bibinfo {author} {\bibfnamefont {S.}~\bibnamefont {Habib}}, \ and\
  \bibinfo {author} {\bibfnamefont {J.~P.}\ \bibnamefont {Paz}},\ }\href
  {\doibase 10.1103/PhysRevLett.70.1187} {\bibfield  {journal} {\bibinfo
  {journal} {Phys. Rev. Lett.}\ }\textbf {\bibinfo {volume} {70}},\ \bibinfo
  {pages} {1187} (\bibinfo {year} {1993})}\BibitemShut {NoStop}%
\bibitem [{\citenamefont {Di{\'o}si}\ and\ \citenamefont
  {Kiefer}(2000)}]{DiosiKieferRobustnessDiffPS}%
  \BibitemOpen
  \bibfield  {author} {\bibinfo {author} {\bibfnamefont {L.}~\bibnamefont
  {Di{\'o}si}}\ and\ \bibinfo {author} {\bibfnamefont {C.}~\bibnamefont
  {Kiefer}},\ }\href@noop {} {\bibfield  {journal} {\bibinfo  {journal} {Phys.
  Rev. Lett.}\ }\textbf {\bibinfo {volume} {85}},\ \bibinfo {pages} {3552}
  (\bibinfo {year} {2000})}\BibitemShut {NoStop}%
\bibitem [{\citenamefont {Eisert}(2004)}]{Eisert2004_ExactDecoToPS}%
  \BibitemOpen
  \bibfield  {author} {\bibinfo {author} {\bibfnamefont {J.}~\bibnamefont
  {Eisert}},\ }\href {\doibase 10.1103/PhysRevLett.92.210401} {\bibfield
  {journal} {\bibinfo  {journal} {Phys. Rev. Lett.}\ }\textbf {\bibinfo
  {volume} {92}},\ \bibinfo {pages} {210401} (\bibinfo {year}
  {2004})}\BibitemShut {NoStop}%
\bibitem [{\citenamefont {Busse}\ and\ \citenamefont
  {Hornberger}(2009)}]{Busse_Hornberger_PS_Nonpertubative_Environment}%
  \BibitemOpen
  \bibfield  {author} {\bibinfo {author} {\bibfnamefont {M.}~\bibnamefont
  {Busse}}\ and\ \bibinfo {author} {\bibfnamefont {K.}~\bibnamefont
  {Hornberger}},\ }\href {http://stacks.iop.org/1751-8121/42/i=36/a=362001}
  {\bibfield  {journal} {\bibinfo  {journal} {J. Phys. A}\ }\textbf {\bibinfo
  {volume} {42}},\ \bibinfo {pages} {362001} (\bibinfo {year}
  {2009})}\BibitemShut {NoStop}%
\bibitem [{\citenamefont {Busse}\ and\ \citenamefont
  {Hornberger}(2010)}]{Busse_Hornberger_PS_CollDeco}%
  \BibitemOpen
  \bibfield  {author} {\bibinfo {author} {\bibfnamefont {M.}~\bibnamefont
  {Busse}}\ and\ \bibinfo {author} {\bibfnamefont {K.}~\bibnamefont
  {Hornberger}},\ }\href {http://stacks.iop.org/1751-8121/43/i=1/a=015303}
  {\bibfield  {journal} {\bibinfo  {journal} {J. Phys. A}\ }\textbf {\bibinfo
  {volume} {43}},\ \bibinfo {pages} {015303} (\bibinfo {year}
  {2010})}\BibitemShut {NoStop}%
\bibitem [{\citenamefont {Lindblad}(1976)}]{Lindblad1976_OnTheGenerators}%
  \BibitemOpen
  \bibfield  {author} {\bibinfo {author} {\bibfnamefont {G.}~\bibnamefont
  {Lindblad}},\ }\href {http://projecteuclid.org/euclid.cmp/1103899849}
  {\bibfield  {journal} {\bibinfo  {journal} {Comm. Math. Phys.}\ }\textbf
  {\bibinfo {volume} {48}},\ \bibinfo {pages} {119} (\bibinfo {year}
  {1976})}\BibitemShut {NoStop}%
\bibitem [{\citenamefont {Gorini}\ \emph {et~al.}(1976)\citenamefont {Gorini},
  \citenamefont {Kossakowski},\ and\ \citenamefont
  {Sudarshan}}]{GoriniKossakowskiSudarshan1976}%
  \BibitemOpen
  \bibfield  {author} {\bibinfo {author} {\bibfnamefont {V.}~\bibnamefont
  {Gorini}}, \bibinfo {author} {\bibfnamefont {A.}~\bibnamefont {Kossakowski}},
  \ and\ \bibinfo {author} {\bibfnamefont {E.~C.~G.}\ \bibnamefont
  {Sudarshan}},\ }\href {\doibase 10.1063/1.522979} {\bibfield  {journal}
  {\bibinfo  {journal} {J. Math. Phys.}\ }\textbf {\bibinfo {volume} {17}},\
  \bibinfo {pages} {821} (\bibinfo {year} {1976})}\BibitemShut {NoStop}%
\bibitem [{\citenamefont {Caldeira}\ and\ \citenamefont
  {Leggett}(1983)}]{Caldeira_Leggett_PathIntegralQBM}%
  \BibitemOpen
  \bibfield  {author} {\bibinfo {author} {\bibfnamefont {A.}~\bibnamefont
  {Caldeira}}\ and\ \bibinfo {author} {\bibfnamefont {A.}~\bibnamefont
  {Leggett}},\ }\href {\doibase 10.1016/0378-4371(83)90013-4} {\bibfield
  {journal} {\bibinfo  {journal} {Physica A}\ }\textbf {\bibinfo {volume}
  {121}},\ \bibinfo {pages} {587} (\bibinfo {year} {1983})}\BibitemShut
  {NoStop}%
\bibitem [{\citenamefont {Unruh}\ and\ \citenamefont
  {Zurek}(1989)}]{UnruhZurek1989_ReductionWavePacketInQBM}%
  \BibitemOpen
  \bibfield  {author} {\bibinfo {author} {\bibfnamefont {W.~G.}\ \bibnamefont
  {Unruh}}\ and\ \bibinfo {author} {\bibfnamefont {W.~H.}\ \bibnamefont
  {Zurek}},\ }\href {\doibase 10.1103/PhysRevD.40.1071} {\bibfield  {journal}
  {\bibinfo  {journal} {Phys. Rev. D}\ }\textbf {\bibinfo {volume} {40}},\
  \bibinfo {pages} {1071} (\bibinfo {year} {1989})}\BibitemShut {NoStop}%
\bibitem [{\citenamefont {Strunz}\ \emph {et~al.}(2003)\citenamefont {Strunz},
  \citenamefont {Haake},\ and\ \citenamefont
  {Braun}}]{StrunzHaakeBraun2003_UnivDecForMacroSuperpos}%
  \BibitemOpen
  \bibfield  {author} {\bibinfo {author} {\bibfnamefont {W.~T.}\ \bibnamefont
  {Strunz}}, \bibinfo {author} {\bibfnamefont {F.}~\bibnamefont {Haake}}, \
  and\ \bibinfo {author} {\bibfnamefont {D.}~\bibnamefont {Braun}},\ }\href
  {\doibase 10.1103/PhysRevA.67.022101} {\bibfield  {journal} {\bibinfo
  {journal} {Phys. Rev. A}\ }\textbf {\bibinfo {volume} {67}},\ \bibinfo
  {pages} {022101} (\bibinfo {year} {2003})}\BibitemShut {NoStop}%
\bibitem [{\citenamefont {{H{\"a}nggi}}\ and\ \citenamefont
  {{Ingold}}(2005)}]{HaenggiIngold2005_FundamentalAspectsQBM}%
  \BibitemOpen
  \bibfield  {author} {\bibinfo {author} {\bibfnamefont {P.}~\bibnamefont
  {{H{\"a}nggi}}}\ and\ \bibinfo {author} {\bibfnamefont {G.-L.}\ \bibnamefont
  {{Ingold}}},\ }\href {\doibase 10.1063/1.1853631} {\bibfield  {journal}
  {\bibinfo  {journal} {Chaos}\ }\textbf {\bibinfo {volume} {15}},\ \bibinfo
  {pages} {026105} (\bibinfo {year} {2005})}\BibitemShut {NoStop}%
\bibitem [{\citenamefont {Strunz}\ \emph {et~al.}(1999)\citenamefont {Strunz},
  \citenamefont {Di{\'o}si}, \citenamefont {Gisin},\ and\ \citenamefont
  {Yu}}]{StrunzDiosiGisin1999_QTForBM}%
  \BibitemOpen
  \bibfield  {author} {\bibinfo {author} {\bibfnamefont {W.~T.}\ \bibnamefont
  {Strunz}}, \bibinfo {author} {\bibfnamefont {L.}~\bibnamefont {Di{\'o}si}},
  \bibinfo {author} {\bibfnamefont {N.}~\bibnamefont {Gisin}}, \ and\ \bibinfo
  {author} {\bibfnamefont {T.}~\bibnamefont {Yu}},\ }\href {\doibase
  10.1103/PhysRevLett.83.4909} {\bibfield  {journal} {\bibinfo  {journal}
  {Phys. Rev. Lett.}\ }\textbf {\bibinfo {volume} {83}},\ \bibinfo {pages}
  {4909} (\bibinfo {year} {1999})}\BibitemShut {NoStop}%
\bibitem [{\citenamefont {Di{\'o}si}(1993)}]{Diosi1993_HighTempMarkovEqForQBM}%
  \BibitemOpen
  \bibfield  {author} {\bibinfo {author} {\bibfnamefont {L.}~\bibnamefont
  {Di{\'o}si}},\ }\href {http://stacks.iop.org/0295-5075/22/i=1/a=001}
  {\bibfield  {journal} {\bibinfo  {journal} {Europhys. Lett.}\ }\textbf
  {\bibinfo {volume} {22}},\ \bibinfo {pages} {1} (\bibinfo {year}
  {1993})}\BibitemShut {NoStop}%
\bibitem [{\citenamefont {Petruccione}\ and\ \citenamefont
  {Vacchini}(2005)}]{PetruccioneVacchini2005_QDescrOfBrownianMotion}%
  \BibitemOpen
  \bibfield  {author} {\bibinfo {author} {\bibfnamefont {F.}~\bibnamefont
  {Petruccione}}\ and\ \bibinfo {author} {\bibfnamefont {B.}~\bibnamefont
  {Vacchini}},\ }\href {\doibase 10.1103/PhysRevE.71.046134} {\bibfield
  {journal} {\bibinfo  {journal} {Phys. Rev. E}\ }\textbf {\bibinfo {volume}
  {71}},\ \bibinfo {pages} {046134} (\bibinfo {year} {2005})}\BibitemShut
  {NoStop}%
\bibitem [{\citenamefont {Vacchini}\ and\ \citenamefont
  {Hornberger}(2007)}]{Vacchini_Hornberger_RelaxationDyn_QBM}%
  \BibitemOpen
  \bibfield  {author} {\bibinfo {author} {\bibfnamefont {B.}~\bibnamefont
  {Vacchini}}\ and\ \bibinfo {author} {\bibfnamefont {K.}~\bibnamefont
  {Hornberger}},\ }\href {\doibase 10.1140/epjst/e2007-00362-9} {\bibfield
  {journal} {\bibinfo  {journal} {Eur. Phys. J. Spec. Top.}\ }\textbf {\bibinfo
  {volume} {151}},\ \bibinfo {pages} {59} (\bibinfo {year} {2007})}\BibitemShut
  {NoStop}%
\bibitem [{\citenamefont {Vacchini}(2000)}]{Vacchini2000_CompPosQuDiss}%
  \BibitemOpen
  \bibfield  {author} {\bibinfo {author} {\bibfnamefont {B.}~\bibnamefont
  {Vacchini}},\ }\href {\doibase 10.1103/PhysRevLett.84.1374} {\bibfield
  {journal} {\bibinfo  {journal} {Phys. Rev. Lett.}\ }\textbf {\bibinfo
  {volume} {84}},\ \bibinfo {pages} {1374} (\bibinfo {year}
  {2000})}\BibitemShut {NoStop}%
\bibitem [{\citenamefont {Hornberger}(2006)}]{Hornberger2006_MEQUParticleGas}%
  \BibitemOpen
  \bibfield  {author} {\bibinfo {author} {\bibfnamefont {K.}~\bibnamefont
  {Hornberger}},\ }\href {\doibase 10.1103/PhysRevLett.97.060601} {\bibfield
  {journal} {\bibinfo  {journal} {Phys. Rev. Lett.}\ }\textbf {\bibinfo
  {volume} {97}},\ \bibinfo {pages} {060601} (\bibinfo {year}
  {2006})}\BibitemShut {NoStop}%
\bibitem [{\citenamefont {Vacchini}\ and\ \citenamefont
  {Hornberger}(2009)}]{Vacchini_Hornberger_QLBE_Review}%
  \BibitemOpen
  \bibfield  {author} {\bibinfo {author} {\bibfnamefont {B.}~\bibnamefont
  {Vacchini}}\ and\ \bibinfo {author} {\bibfnamefont {K.}~\bibnamefont
  {Hornberger}},\ }\href {\doibase 10.1016/j.physrep.2009.06.001} {\bibfield
  {journal} {\bibinfo  {journal} {Phys. Rep.}\ }\textbf {\bibinfo {volume}
  {478}},\ \bibinfo {pages} {71} (\bibinfo {year} {2009})}\BibitemShut
  {NoStop}%
\bibitem [{\citenamefont {Gallis}\ and\ \citenamefont
  {Fleming}(1990)}]{GallisFleming1990}%
  \BibitemOpen
  \bibfield  {author} {\bibinfo {author} {\bibfnamefont {M.~R.}\ \bibnamefont
  {Gallis}}\ and\ \bibinfo {author} {\bibfnamefont {G.~N.}\ \bibnamefont
  {Fleming}},\ }\href {\doibase 10.1103/PhysRevA.42.38} {\bibfield  {journal}
  {\bibinfo  {journal} {Phys. Rev. A}\ }\textbf {\bibinfo {volume} {42}},\
  \bibinfo {pages} {38} (\bibinfo {year} {1990})}\BibitemShut {NoStop}%
\bibitem [{\citenamefont {Hornberger}\ and\ \citenamefont
  {Sipe}(2003)}]{HornbergerSipe2003_CollDecoReexamined}%
  \BibitemOpen
  \bibfield  {author} {\bibinfo {author} {\bibfnamefont {K.}~\bibnamefont
  {Hornberger}}\ and\ \bibinfo {author} {\bibfnamefont {J.~E.}\ \bibnamefont
  {Sipe}},\ }\href {\doibase 10.1103/PhysRevA.68.012105} {\bibfield  {journal}
  {\bibinfo  {journal} {Phys. Rev. A}\ }\textbf {\bibinfo {volume} {68}},\
  \bibinfo {pages} {012105} (\bibinfo {year} {2003})}\BibitemShut {NoStop}%
\bibitem [{\citenamefont {Breuer}\ and\ \citenamefont
  {Petruccione}(2006)}]{Breuer_Petruccione_OpenQS}%
  \BibitemOpen
  \bibfield  {author} {\bibinfo {author} {\bibfnamefont {H.-P.}\ \bibnamefont
  {Breuer}}\ and\ \bibinfo {author} {\bibfnamefont {F.}~\bibnamefont
  {Petruccione}},\ }\href@noop {} {\emph {\bibinfo {title} {{The Theory of Open
  Quantum Systems}}}}\ (\bibinfo  {publisher} {Oxford University Press},\
  \bibinfo {year} {2006})\BibitemShut {NoStop}%
\bibitem [{\citenamefont {Zurek}(1993)}]{Zurek1993_PredictSieve}%
  \BibitemOpen
  \bibfield  {author} {\bibinfo {author} {\bibfnamefont {W.~H.}\ \bibnamefont
  {Zurek}},\ }\href@noop {} {\bibfield  {journal} {\bibinfo  {journal} {Prog.
  Theor. Phys.}\ }\textbf {\bibinfo {volume} {89}},\ \bibinfo {pages} {281}
  (\bibinfo {year} {1993})}\BibitemShut {NoStop}%
\bibitem [{\citenamefont {Dalvit}\ \emph {et~al.}(2005)\citenamefont {Dalvit},
  \citenamefont {Dziarmaga},\ and\ \citenamefont {Zurek}}]{Zurek_predSieve_Ps}%
  \BibitemOpen
  \bibfield  {author} {\bibinfo {author} {\bibfnamefont {D.~A.~R.}\
  \bibnamefont {Dalvit}}, \bibinfo {author} {\bibfnamefont {J.}~\bibnamefont
  {Dziarmaga}}, \ and\ \bibinfo {author} {\bibfnamefont {W.~H.}\ \bibnamefont
  {Zurek}},\ }\href {\doibase 10.1103/PhysRevA.72.062101} {\bibfield  {journal}
  {\bibinfo  {journal} {Phys. Rev. A}\ }\textbf {\bibinfo {volume} {72}},\
  \bibinfo {pages} {062101} (\bibinfo {year} {2005})}\BibitemShut {NoStop}%
\bibitem [{\citenamefont {Di{\'o}si}(1986)}]{Diosi1986_StochasticPureStateRep}%
  \BibitemOpen
  \bibfield  {author} {\bibinfo {author} {\bibfnamefont {L.}~\bibnamefont
  {Di{\'o}si}},\ }\href {\doibase 10.1016/0375-9601(86)90692-4} {\bibfield
  {journal} {\bibinfo  {journal} {Phys. Lett. A}\ }\textbf {\bibinfo {volume}
  {114}},\ \bibinfo {pages} {451} (\bibinfo {year} {1986})}\BibitemShut
  {NoStop}%
\bibitem [{\citenamefont {Gisin}\ and\ \citenamefont
  {Rigo}(1995)}]{Gisin_Rigo_Nonlin_Ses}%
  \BibitemOpen
  \bibfield  {author} {\bibinfo {author} {\bibfnamefont {N.}~\bibnamefont
  {Gisin}}\ and\ \bibinfo {author} {\bibfnamefont {M.}~\bibnamefont {Rigo}},\
  }\href {http://stacks.iop.org/0305-4470/28/i=24/a=030} {\bibfield  {journal}
  {\bibinfo  {journal} {J. Phys. A}\ }\textbf {\bibinfo {volume} {28}},\
  \bibinfo {pages} {7375} (\bibinfo {year} {1995})}\BibitemShut {NoStop}%
\bibitem [{\citenamefont {Gardiner}\ \emph {et~al.}(1992)\citenamefont
  {Gardiner}, \citenamefont {Parkins},\ and\ \citenamefont
  {Zoller}}]{Gardiner1992_QSDEMethods}%
  \BibitemOpen
  \bibfield  {author} {\bibinfo {author} {\bibfnamefont {C.~W.}\ \bibnamefont
  {Gardiner}}, \bibinfo {author} {\bibfnamefont {A.~S.}\ \bibnamefont
  {Parkins}}, \ and\ \bibinfo {author} {\bibfnamefont {P.}~\bibnamefont
  {Zoller}},\ }\href@noop {} {\bibfield  {journal} {\bibinfo  {journal} {Phys.
  Rev. A}\ }\textbf {\bibinfo {volume} {46}},\ \bibinfo {pages} {4363}
  (\bibinfo {year} {1992})}\BibitemShut {NoStop}%
\bibitem [{\citenamefont {Dalibard}\ \emph {et~al.}(1992)\citenamefont
  {Dalibard}, \citenamefont {Castin},\ and\ \citenamefont
  {M{\o}lmer}}]{DalibardCastinMolmer1992_WaveFunctionApprInQO}%
  \BibitemOpen
  \bibfield  {author} {\bibinfo {author} {\bibfnamefont {J.}~\bibnamefont
  {Dalibard}}, \bibinfo {author} {\bibfnamefont {Y.}~\bibnamefont {Castin}}, \
  and\ \bibinfo {author} {\bibfnamefont {K.}~\bibnamefont {M{\o}lmer}},\ }\href
  {\doibase 10.1103/PhysRevLett.68.580} {\bibfield  {journal} {\bibinfo
  {journal} {Phys. Rev. Lett.}\ }\textbf {\bibinfo {volume} {68}},\ \bibinfo
  {pages} {580} (\bibinfo {year} {1992})}\BibitemShut {NoStop}%
\bibitem [{\citenamefont {M{\o}lmer}\ \emph {et~al.}(1993)\citenamefont
  {M{\o}lmer}, \citenamefont {Castin},\ and\ \citenamefont
  {Dalibard}}]{Molmer1993_MonteCarloMethodsQO}%
  \BibitemOpen
  \bibfield  {author} {\bibinfo {author} {\bibfnamefont {K.}~\bibnamefont
  {M{\o}lmer}}, \bibinfo {author} {\bibfnamefont {Y.}~\bibnamefont {Castin}}, \
  and\ \bibinfo {author} {\bibfnamefont {J.}~\bibnamefont {Dalibard}},\ }\href
  {\doibase 10.1364/JOSAB.10.000524} {\bibfield  {journal} {\bibinfo  {journal}
  {J. Opt. Soc. Am. B}\ }\textbf {\bibinfo {volume} {10}},\ \bibinfo {pages}
  {524} (\bibinfo {year} {1993})}\BibitemShut {NoStop}%
\bibitem [{\citenamefont {Rigo}\ and\ \citenamefont
  {Gisin}(1996)}]{RigoGisin1996_UnravMEEmergenceClass}%
  \BibitemOpen
  \bibfield  {author} {\bibinfo {author} {\bibfnamefont {M.}~\bibnamefont
  {Rigo}}\ and\ \bibinfo {author} {\bibfnamefont {N.}~\bibnamefont {Gisin}},\
  }\href {http://stacks.iop.org/1355-5111/8/i=1/a=018} {\bibfield  {journal}
  {\bibinfo  {journal} {Quantum Semiclass. Opt.}\ }\textbf {\bibinfo {volume}
  {8}},\ \bibinfo {pages} {255} (\bibinfo {year} {1996})}\BibitemShut {NoStop}%
\bibitem [{\citenamefont {Gisin}(1984)}]{Gisin1984_QuMeasStochProc}%
  \BibitemOpen
  \bibfield  {author} {\bibinfo {author} {\bibfnamefont {N.}~\bibnamefont
  {Gisin}},\ }\href {\doibase 10.1103/PhysRevLett.52.1657} {\bibfield
  {journal} {\bibinfo  {journal} {Phys. Rev. Lett.}\ }\textbf {\bibinfo
  {volume} {52}},\ \bibinfo {pages} {1657} (\bibinfo {year}
  {1984})}\BibitemShut {NoStop}%
\bibitem [{\citenamefont
  {Diosi}(1988)}]{Diosi1988_QuantStochProcModelStateVectorRed}%
  \BibitemOpen
  \bibfield  {author} {\bibinfo {author} {\bibfnamefont {L.}~\bibnamefont
  {Diosi}},\ }\href {\doibase 10.1088/0305-4470/21/13/013} {\bibfield
  {journal} {\bibinfo  {journal} {J. Phys. A}\ }\textbf {\bibinfo {volume}
  {21}},\ \bibinfo {pages} {2885} (\bibinfo {year} {1988})}\BibitemShut
  {NoStop}%
\bibitem [{\citenamefont {Ghirardi}\ \emph {et~al.}(1990)\citenamefont
  {Ghirardi}, \citenamefont {Pearle},\ and\ \citenamefont
  {Rimini}}]{GhirardiPearleRimini1990_MarkovProcessesHilbertSpaceandCSL}%
  \BibitemOpen
  \bibfield  {author} {\bibinfo {author} {\bibfnamefont {G.~C.}\ \bibnamefont
  {Ghirardi}}, \bibinfo {author} {\bibfnamefont {P.}~\bibnamefont {Pearle}}, \
  and\ \bibinfo {author} {\bibfnamefont {A.}~\bibnamefont {Rimini}},\ }\href
  {\doibase 10.1103/PhysRevA.42.78} {\bibfield  {journal} {\bibinfo  {journal}
  {Phys. Rev. A}\ }\textbf {\bibinfo {volume} {42}},\ \bibinfo {pages} {78}
  (\bibinfo {year} {1990})}\BibitemShut {NoStop}%
\bibitem [{\citenamefont {Gisin}\ and\ \citenamefont
  {Percival}(1992)}]{GisinPercival1992_QSDAppliedToOpenSys}%
  \BibitemOpen
  \bibfield  {author} {\bibinfo {author} {\bibfnamefont {N.}~\bibnamefont
  {Gisin}}\ and\ \bibinfo {author} {\bibfnamefont {I.~C.}\ \bibnamefont
  {Percival}},\ }\href {http://stacks.iop.org/0305-4470/25/i=21/a=023}
  {\bibfield  {journal} {\bibinfo  {journal} {J. Phys. A}\ }\textbf {\bibinfo
  {volume} {25}},\ \bibinfo {pages} {5677} (\bibinfo {year}
  {1992})}\BibitemShut {NoStop}%
\bibitem [{\citenamefont {Fleck}\ \emph {et~al.}(1976)\citenamefont {Fleck},
  \citenamefont {Morris},\ and\ \citenamefont {Feit}}]{Fleck1976_FFTMethod}%
  \BibitemOpen
  \bibfield  {author} {\bibinfo {author} {\bibfnamefont {J.~J.}\ \bibnamefont
  {Fleck}}, \bibinfo {author} {\bibfnamefont {J.}~\bibnamefont {Morris}}, \
  and\ \bibinfo {author} {\bibfnamefont {M.}~\bibnamefont {Feit}},\ }\href
  {\doibase 10.1007/BF00896333} {\bibfield  {journal} {\bibinfo  {journal}
  {Appl. Phys.}\ }\textbf {\bibinfo {volume} {10}},\ \bibinfo {pages} {129}
  (\bibinfo {year} {1976})}\BibitemShut {NoStop}%
\bibitem [{\citenamefont {Feit}\ \emph {et~al.}(1982)\citenamefont {Feit},
  \citenamefont {Jr.},\ and\ \citenamefont
  {Steiger}}]{Feit1982_SplitOpFFTSchroedeq}%
  \BibitemOpen
  \bibfield  {author} {\bibinfo {author} {\bibfnamefont {M.}~\bibnamefont
  {Feit}}, \bibinfo {author} {\bibfnamefont {J.~F.}\ \bibnamefont {Jr.}}, \
  and\ \bibinfo {author} {\bibfnamefont {A.}~\bibnamefont {Steiger}},\ }\href
  {\doibase 10.1016/0021-9991(82)90091-2} {\bibfield  {journal} {\bibinfo
  {journal} {J. Comput. Phys.}\ }\textbf {\bibinfo {volume} {47}},\ \bibinfo
  {pages} {412} (\bibinfo {year} {1982})}\BibitemShut {NoStop}%
\bibitem [{\citenamefont {Press}\ \emph {et~al.}(2007)\citenamefont {Press},
  \citenamefont {Teukolsky}, \citenamefont {Vetterling},\ and\ \citenamefont
  {Flannery}}]{Press2007_NumRec}%
  \BibitemOpen
  \bibfield  {author} {\bibinfo {author} {\bibfnamefont {W.~H.}\ \bibnamefont
  {Press}}, \bibinfo {author} {\bibfnamefont {S.~A.}\ \bibnamefont
  {Teukolsky}}, \bibinfo {author} {\bibfnamefont {W.~T.}\ \bibnamefont
  {Vetterling}}, \ and\ \bibinfo {author} {\bibfnamefont {B.~P.}\ \bibnamefont
  {Flannery}},\ }\href@noop {} {\emph {\bibinfo {title} {{Numerical Recipes 3rd
  Edition: The Art of Scientific Computing}}}},\ \bibinfo {edition} {3rd}\ ed.\
  (\bibinfo  {publisher} {Cambridge University Press},\ \bibinfo {address} {New
  York, NY, USA},\ \bibinfo {year} {2007})\BibitemShut {NoStop}%
\bibitem [{\citenamefont {Gardiner}(2009)}]{gardiner2004handbook}%
  \BibitemOpen
  \bibfield  {author} {\bibinfo {author} {\bibfnamefont {C.~W.}\ \bibnamefont
  {Gardiner}},\ }\href@noop {} {\emph {\bibinfo {title} {{Handbook of
  stochastic methods for physics, chemistry and the natural sciences}}}},\
  \bibinfo {edition} {4th}\ ed.\ (\bibinfo  {publisher} {Springer-Verlag
  Berlin},\ \bibinfo {year} {2009})\BibitemShut {NoStop}%
\bibitem [{\citenamefont {Risken}(1989)}]{Risken1989_FPE}%
  \BibitemOpen
  \bibfield  {author} {\bibinfo {author} {\bibfnamefont {H.}~\bibnamefont
  {Risken}},\ }\href@noop {} {\emph {\bibinfo {title} {{The Fokker-Planck
  Equation: Methods of Solutions and Applications}}}},\ \bibinfo {edition}
  {2nd}\ ed.\ (\bibinfo  {publisher} {Springer-Verlag Berlin},\ \bibinfo {year}
  {1989})\BibitemShut {NoStop}%
\end{thebibliography}

%

\end{document}